\documentclass{article}
\usepackage{jcappub}
\usepackage[utf8]{inputenc}
\usepackage[dvipsnames]{xcolor}
\usepackage{comment}
\usepackage[english]{babel}
\usepackage{amsmath}
\usepackage{graphicx,subcaption}
\graphicspath{ {images/} }
\usepackage{float}
\usepackage{amssymb}
\usepackage{caption}
\usepackage{subcaption}
\usepackage{adjustbox}
\usepackage{tabularx}
\usepackage{multicol}
\usepackage{multirow}
\usepackage{slashed}
\usepackage{feynmp-auto}
\usepackage{comment}

\usepackage[tableposition=below]{caption}
\usepackage{longtable}
\usepackage{color}			
\usepackage{verbatim}

\newcommand{\ed}{\mathcal{E}}

\newcommand{\geff}{\mathfrak{g}}

\newcommand{\mkin}{\tilde{m}}

\newcommand{\mr}{r_\mathrm{m}}

\newcommand{\cp}{\mu}

\newcommand{\ce}{\epsilon}

\newcommand{\beq}{\begin{eqnarray}}
\newcommand{\eeq}{\end{eqnarray}}

\newcommand{\ngg}{\mathrel{/\!\!\!\!\!\!\gg}}

\numberwithin{equation}{section} 

\title{Structure of Stellar Remnants with Coupling to a Light Scalar}
\author{Christina Gao}
\author{and Albert Stebbins}
\affiliation{Theory Department, Fermi National Accelerator Laboratory, Batavia, IL 60510, USA}

\emailAdd{yanggao@fnal.gov}
\emailAdd{stebbins@fnal.gov}

\abstract{
In this paper we study how a Yukawa coupling of the Standard Model fermions to a light scalar field effects the stellar structure of cold stellar remnants such as neutron stars.  We elucidate the stellar structure phenomenology using a simple model of a massive scalar coupled to a single dominant fermion with no other interactions.  For a broad scalar mass range ($10^{-10}\,\mathrm{eV}\ll m_\phi\ll10^3\,\mathrm{eV}$ for neutron stars) we show that the equation-of-state and stellar structure depends only the effective coupling $\geff=\frac{g_f\,m_f}{m_\phi}$, where $g_f$ is the Yukawa coupling, $m_f$ the fermion mass and $m_\phi$ is the scalar kinematic mass at nuclear densities.  If $\geff>\mathcal{O}(1)$ the Yukawa coupled matter exhibits various anomalous behaviors including hydrodynamic instability, negative pressure, distinct phases (soft and hard) of matter with sharp phase boundaries between them and with vacuum.  These anomalies can lead to stars consisting of only soft, only hard or hybrid of soft and hard matter.  These stars can have varying signs of the slope of the mass-radius relation, anomalously large and small masses, gaps in allowed radii, multiple radii for the same mass, very thin crusts and radiate anomalously large amounts of energy when they form (in the form of neutrinos for neutron stars).  To the extent that these anomalies have not and/or will not be observed limits the effective coupling to $\geff<\mathcal{O}(1)$.  We argue this phenomenology is generic to realistic models of stars with Yukawa coupled matter.
}

\begin{document}

\maketitle

\section{Introduction}

A light scalar field is one of the simplest extension to the Standard Model (SM) but is only one of the plethora of proposed extensions referred to collectively as Beyond the Standard Model (BSM) physics.  A scalar can act as a ``mediator'' between the SM and the ``dark sector'' of particles which is believed to be associated with a putative dark matter particle, i.e. it couples directly to SM and dark sector particles.  The direct SM coupling could make a mediator easier to detect than other dark sector particles.  Such a mediator will also give a non-SM (BSM) coupling of SM matter with itself which is the topic of this paper.   We make no reference to a dark sector or to dark matter.  Here we consider the effect of a linear scalar field coupling to SM particles on the structure of stellar remnants.

Neutron stars provide a unique astrophysical laboratory for both SM and BSM physics in the combination of properties
\begin{enumerate}
\item like many astronomical objects, gravity determines their structure, which provides a probe of BSM interactions of matter at the gravitational strength in addition to or in place of the standard gravity (GR / general relativity),
\item their small size ($\sim10\,$km) provides a probe of these interactions on smaller length scales than other astronomical objects
\item their structure on these small scales is well constrained by a variety of very different types of observations,
\item they consist of bulk quantities of nuclear density matter (the highest density in the present universe) providing a probe of SM and BSM interactions at these densities as well as their density dependence,
\item their SM equation of state (EoS) is (ideally) simple with essentially no free parameters providing a foil against which to test BSM effects.  
\end{enumerate}
While there may be no free parameters in the SM EoS, there is currently a great deal of uncertainty in the EoS of cold matter at the nuclear densities due to difficulties in accounting for the SM hadronic interactions.  The uncertainty in the SM EoS leads to the uncertainty in neutron star structure and thus limits observational sensitivities to BSM effects from the observations of neutron star structure.

The simple EoS property derives from a short cooling time ($\sim10\,$sec by neutrinos) allowing the bulk\footnote{The state of the surface (crust) of a neutron star may be impacted by the exterior environment and history.} to cool well below the Fermi energy of its constituent particles and therefore $T=0$ is a good approximation.  Before cooling, the bulk of the stellar matter was hot and dense enough to approach nuclear statistical equilibrium (NSE) which means that the composition is known and described by a single baryon chemical potential\footnote{Conductivity by $e^-$, $p^+$, etc. leads to a negligible charge separation.}.  At nuclear densities, neutron star consists primarily of neutrons.  In a given SM or BSM context, a neutron star is like a black hole in that it is well described by only its mass and angular momentum.  Unlike black hole stellar remnants, the observed and expected rotation rate of neutron stars is small enough to have little effect on their structure.  While black holes have ``no hair", neutron stars can and do; most importantly the embedded magnetic fields. However these fields are weak enough to have even less effect on the structure of the star than angular momentum.  To $\lesssim10\%$ accuracy, mass is the only parameter of neutron star structure, although angular momentum\footnote{Neutron stars used to constrain BSM physics would have to have known rotation rates.} and to a lesser extent magnetic fields would become important if we could determine its structure with greater precision.  The stars are thus nearly spherically symmetric and their structure is given by their density profile or $M(R)$.  This paper focuses on determining $M(R)$ when normal matter is coupled to a scalar field.

White dwarfs share much of the simplicity of neutron stars.  While white dwarfs are not in NSE and their chemical composition will vary, there is relatively little composition dependence of their EoS.  In \S\ref{sec:WhiteDwarfs} we briefly discuss how our findings for neutrons stars apply almost directly to white dwarfs.
 
\subsection{Overview}
\label{sec:Overview}

In this paper we examine how the existence of a Yukawa coupling to a BSM scalar field will effect the structure of stellar remnants.  It is not assumed that this scalar is present in any significant abundance in the universe, say, as dark matter. The paper is organized in the following way.  In \S\ref{sec:YukawaCoupling} we describe the coupling under consideration and the level of approximation with which we treat it.  In \S\ref{sec:StellarStructure} we review stellar structure and the ingredient needed to determine this structure which is the equation of state (EoS).  In \S\ref{sec:IGY} we formulate the EoS including a scalar coupled to an ideal gas of free fermions.  Specifically in \S\ref{sec:polarization} we show how the scalar is polarized to non-zero field values in the presence of fermions. In \S\ref{sec:HeavyLightUltralight} we identify three regimes for computation of this polarization, {\it heavy}, {\it light} and {\it ultralight} which depends on the mass of the scalar and the density of fermions.  Specializing to the most interesting light scalar regime in \S\ref{sec:IGYEoS} we give analytic expressions for the ideal gas +Yukawa, or IGY EoS which for a single fermion depends only on the {\it effective coupling} $\geff_f=\frac{g_f\,m_f}{m_\phi}$ ($g_f$ is the Yukawa coupling, $m_f$ the fermion mass and $m_\phi$ the scalar mass). In \S\ref{sec:anomalies} we highlight anomalies in the IGY EoS which appear when $\geff_f$ exceeds a {\it critical} value.  These anomalies lead to different matter {\it phases} depending on the matter density:  a range of densities {\it forbidden} in stars separates a low density {\it soft} matter phase from a higher density {\it hard} matter phase.  The soft phase can exist in a {\it metastable} state and all low density matter may be metastable to conversion to hard matter if $\geff_f$ is large enough.   For a small range of $\geff_f$ a stable phase boundary can exist between hard and soft matter (\S\ref{sec:SoftHardInterface}) and for larger $\geff_f$ a stable boundary between hard matter and vacuum (\S\ref{sec:VacuumHardInterface}).   {\it  IGY stars}, which are spherical hydrostatic configurations of matter with a single fermion (the neutron) IGY EoS provide a prototype for neutron stars with a Yukawa coupling between a scalar and baryons.  This simple model is only considered a prototype as it has a number of deficiencies listed in \S\ref{sec:IGYdeficiencies}. This phenomenology is described in \S\ref{sec:IGYstellarStructure} showing their masses and radii in \S\ref{sec:MassRadius}-\S\ref{sec:MaximalStars} and selected density profiles in \S\ref{sec:DensityProfiles}.  Above the critical coupling stars may consist purely of hard matter or of soft matter or a hybrid of the two; and some stars may be metastable.  Very high density but very low mass hard matter configurations, bound by the scalar field and not by gravity, exist which we call {\it hard drops} (\S\ref{sec:HardDrops}) which could constitute the dark matter.  In \S\ref{sec:BindingEnergy} we explore the energetics of collapse to an IGY star and the implications for the energy radiated during formation.  In \S\ref{sec:IGYStarAnomalies} we list some of the anomalous features of the IGY stellar structure relative to that of SM stars when $\geff_f$  is large enough:  the mass may increase or decrease with radius; the maximum mass stable star may or may not have the minimum radius;  there may be a gap in allowed stellar radii; there are stable configurations with nuclear densities but extremely small radii and masses; stars much larger in size and mass than the SM neutron stars are possible; the crust of an IGY star may be very thin; neutrino bursts from IGY stars may be very large.  We argue in \S\ref{sec:IGYgenerality} that the qualitative phenomenology found for IGY stars will be common to most models of stellar remnants with matter coupled to scalars.   In \S\ref{sec:WhiteDwarfs} we illustrate this point with the example of white dwarfs.  In \S\ref{sec:Implications} we discuss implications of our results for detecting or excluding a Yukawa coupling from observations of neutron stars and white dwarfs: $\geff_f<\mathcal{O}(1)$ if no anomalous behavior is observed where $f$ is fermion which dominates the pressure.   More realistic modeling of Yukawa-coupled neutron stars and white dwarfs and comparison to data are left to future works which will set bounds on Yukawa couplings from astronomical observations of stellar structure and compare these with bounds obtained by other methods.

The EoS phenomenology developed here has much in common with that developed for mass-varying neutrinos \cite{AfshordiZaldarriagaKohri2005,EstebanSalvado2021}.

\subsection{Conventions and Notation}

Neutron stars are mildly relativistic so our calculations should account for space-time curvature.  We assume Einstein metric gravity and use the $\{+,-,-,-\}$ metric signature convention, Einstein summation convention and ${}_{;\mu}$ for covariant derivatives.  We will use HEP natural units where $\hbar=c=1$ except where they appear explicitly.

\section{A Scalar Linearly Coupled To Matter}
\label{sec:YukawaCoupling}

We posit the existence of a light BSM real scalar field $\phi$ which has a linear coupling to quarks, leptons, gluons and/or photons\footnote{Here  for simplicity the coupling is assumed diagonal between different quarks and gluons.}  
\beq
\mathcal{L}\supset\left(\sum_q g_q\,\bar{\psi}_q\,\psi_q+\sum_\ell g_\ell\,\bar{\psi}_\ell\,\psi_\ell
+g_\mathrm{g}\,\sum_a G^{\mu\nu}_a\,G^a_{\mu\nu}+g_\mathrm\gamma\,F^{\mu\nu}\,F_{\mu\nu}\right)\,  \phi
\label{eq:linearCoupling}
\eeq
where $q$ / $\ell$ / $a$ sum over the different quarks / leptons / gluons,  $\psi_x$ are the fermion spinors and $G^a_{\mu\nu}$ / $F_{\mu\nu}$ are the gluon / electromagnetic field tensors.  $g_q$, $g_\ell$ are dimensionless Yukawa coupling constants.  $g_\mathrm{g}$ and $g_\gamma$ are scalar couplings with dimension mass$^{-1}$. We define $\phi$ such that $\phi=0$ gives the minimum energy in the absence of matter.  We assume minimal coupling of $\phi$ to gravity, e.g. no significant $\phi\,R$ term in $\mathcal{L}$.  The couplings are taken as free parameters but clearly must be extremely small for $\phi$ to have escaped detection.   The main goal of this paper is to explore how such couplings modify stellar structure 
if such a $\phi$ exists.

If $\phi$ were a constant then one could absorb the $g$'s by suitable field redefinitions and adjustment of mass matrices and gauge couplings.  However we show that the value of $\phi$ will vary according to the macroscopic environment (``finite density'' effects) resulting in environment dependent SM particle parameters and properties which differ from the near vacuum values measured in the laboratory.  This is analogous to the Higgs mechanism in SM electro-weak symmetry breaking in which particles attain mass.  Here however there is no phase transition, the environment dependence is gradual and in static stars purely spatial.  Environment dependence plays a key role in modified gravity / 5th-force models through a chameleon mechanism which is analogous but different to that considered here. $\phi$ can contribute to a long range (5th) force but we will focus on parameter regimes where such forces do not extend significantly beyond the stellar surface.

Astrophysical phenomena are usually formulated in terms of a few of  the fundamental particles of eq.~\ref{eq:linearCoupling} plus other strongly or weakly\footnote{meaning the binding energy much less than the mass of the constituent particles} bound composites of the other elementary constituents.  These composite particles will inherit a coupling to $\phi$ through their constituents.  To 1st order in $\phi$ the form of the effective interaction with $\phi$ for composite particles is dictated by Lorentz invariance which for unpolarized fermions is
\beq
\mathcal{L}_\mathrm{eff}\supset\phi\,\sum_f\,g_f \,\bar{\psi}_f\,\psi_f
\label{eq:effectiveCoupling}
\eeq
where $\psi_f$ is the spinor of fermion $f$.  In an astrophysical context particles are rarely significantly polarized, and even if not, a spin coupling is likely a subdominant effect.\footnote{
A non-zero $\bar{\psi}_f\,\psi_f$ is analogous to a bulk space-charge density in electromagnetism without oppositely signed charges to shield long range fields.  Bulk space charges leads to much larger field strengths than spin couplings which, even for fully polarized materials, contributes only surface and gradient terms as, say, in magnetic materials.}
For weakly bound composites such as multi-nucleon nuclei and atoms the coupling constant will approximately be the sum of the couplings of the constituents, e.g. for nuclei $g_{(Z,A)}\approx Z\,g_\mathrm{p}+(A-Z)\,g_\mathrm{n}$ and neutral atoms add $+Z\,g_\mathrm{e}$.   The three couplings, $g_\mathrm{p}$, $g_\mathrm{n}$, $g_\mathrm{e}$, are the most important for stellar structure. 

In principle $g_\mathrm{p}$ and $g_\mathrm{n}$ are related to the more fundamental couplings of eq.~\ref{eq:linearCoupling}.  However hadrons are strongly bound composites of quarks and gluons with no simple relation between $g_h$ ($h\in$ nucleons, mesons, etc.) and the more fundamental couplings.  Determining this relation would require knowledge of $\phi$-dependent hadronic structure.   This structure has not yet been fully understood even in the SM and is outside the scope of this paper.   The relative size of $g_h$ depends on the relative contribution of quark and gluon couplings which adds further uncertainty.  Nuclear interactions which are important in neutron stars are mediated by mesons which are mostly pseudo-scalars and cannot have linear couplings, e.g. $\pi^0\,\phi$, as in eq.~\ref{eq:effectiveCoupling} though they can have other couplings e.g. quadratic $\pi^0\,\pi^0\,\phi$.  We leave discussion of nuclear interactions to a follow-on paper.

\subsection{Effective Mean Field Theory}

The $\phi$ field as a force carrier whose range is given by the mass of $\phi$ which if small will coherently interact with a very large number, say $N$, of SM particles. This effectively boosts the interaction energy by a factor of $N$ over single particle incoherent interactions.  Small mass or large $N$ can lead to interaction energy between $\phi$ and the SM matter exceeding SM forces even for extremely small couplings.  SM forces do not have large coherent boosts either because of the large mass of the force carriers in the case of nuclear interactions or charge screening in the case of electromagnetism.  Due to the finite range of the force the boost factor $\sim N$ increases as the density so manifestations of $\phi$ can dominate at the large densities of white dwarfs or neutron stars while being very small at laboratory densities.  In vacuum, $\phi$ may fluctuate around zero, but in a high density environment with large $N$, the lowest energy state will have very large $\phi$ values, thus in a well cooled and equilibrated star, these non-zero values, $\tilde{\phi}\equiv\langle\phi\rangle$, will be attained.  Stars are not uniform and  $\tilde{\phi}$ will vary within the star.  We treat the SM particles in the same mean field approximation supposing they have reached some local equilibrium.  This mean field approach does not reveal certain phenomena such as $\phi$ production from inelastic SM particle scatterings, but these processes are suppressed by $N$ and in most cases of interest are negligible.  Scattering production of $\phi$'s can also be highly suppressed by Fermi blocking. 

 The effective action of the system under study is
\beq
S=\int d^4x^\mu\,\sqrt{-\mathrm{g}}\,\mathcal{L}_\mathrm{eff} \qquad
\mathcal{L}_\mathrm{eff} =\mathcal{L}_\mathrm{SM}+\tilde{g}\,\tilde{\phi}\,\tilde{n}+\mathcal{L}_{\tilde{\phi}}
\label{eq:action}
\eeq 
where $\mathcal{L}_\mathrm{SM}$ is the standard model Lagrangian density,
$\mathcal{L}_{\tilde{\phi}}=\frac{\tilde{\phi}^{;\mu}\,\tilde{\phi}_{;\mu}}{2}-V(\tilde{\phi})$, and
$\mathrm{g}$ is the determinant of the metric tensor (not to be confused with $\tilde{g}$ which is an unspecified weighted average of the couplings previously described), $\tilde{n}$ is the corresponding sum of mean field values, $\langle\bar{\psi}\,\psi\rangle$'s.  When a single free fermion species, say neutrons, dominates, one should choose $\tilde{g}=g_\mathrm{n}$ and $\tilde{n}=\langle\bar{\psi}_\mathrm{n}\,\psi_\mathrm{n}\rangle$.

Note that $\tilde{n}$ has dimensions of particle number density and is related to the particle number density of nucleons as we shall show.  Unlike the frame dependent volume number density, $\tilde{n}$ is a Lorentz scalar and does not invoke any reference frame (4-velocity $u^\mu$) such as the center of momentum frame of a fluid.  Therefore, there is  no possibility of a conserved current ($(\tilde{n}\,u^\mu)_{;\mu}=0$) even when particle numbers are conserved.

\section{Stellar Structure Equations}
\label{sec:StellarStructure}

The structure of a star is given by the stress-energy $T_{\mu\nu}$. The Hilbert stress-energy\footnote{which may be generalized to include spinor fields}, $T_{\mu\nu}=2\,\frac{\delta\mathcal{L}}{\delta \mathrm{g}^{\mu\nu}}-\mathrm{g}_{\mu\nu}\,\mathcal{L}$, may be decomposed as
\beq
T_{\mu\nu}=T^\mathrm{SM}_{\mu\nu}+
2\,\tilde{g}\,\tilde{\phi}\,\frac{\delta\tilde{n}}{\delta \mathrm{g}^{\mu\nu}}
-\tilde{g}\,\tilde{\phi}\,\tilde{n}\,\mathrm{g}_{\mu\nu}+T^{\tilde{\phi}}_{\mu\nu}\ .
\label{eq:StressEnergy}
\eeq
Spherical static stars have time-translation and 3-D rotational isometries which dictate the form of the $\mathrm{g}_{\mu\nu}$ and $T_{\mu\nu}$, which may be written in spherical polar coordinates $\{t,r,\theta,\phi\}$ as
\begin{eqnarray}
\mathrm{g}_{\mu\nu}&=&\mathrm{diag}\left(e^{2\Phi(r)},\frac{-1}{1-\frac{2\,G\,M(r)}{r}},-r^2,-r^2\,\sin^2\theta\right)  \nonumber \\
{T^\mu}_\nu&=&\mathrm{diag}\left(\ed(r),-P_\parallel(r),-P_\perp(r),-P_\perp(r)\right)
\label{eq:metricStressEnergy}
\end{eqnarray}
where  $r$ is the circumferential radius, $\ed$,  $P_\parallel$ and $P_\perp$  are respectively the energy density, radial and tangential pressure in the stellar rest frame. The two metric functions $M(r)$ and $\Phi(r)$ were chosen to be similar to their Newtonian analogs.  Einstein's equations may be written as
\begin{eqnarray}
\label{eq:EinsteinEquations}
M'(r)&=&4\pi r^2\ed(r) \\
\Phi'(r)&=&\frac{G\,(M(r)+4\,\pi\,r^3 P_\parallel(r))}{r^2\,\left(1-\frac{2\,G\,M(r)}{r}\right)}
=\frac{2}{r}\frac{P_\perp(r)-P_\parallel(r)}{\ed(r)+P_\parallel(r)}-\frac{P_\parallel'(r)}{\ed(r)+P_\parallel(r)} \nonumber
\end{eqnarray}
where $f'(r)\equiv\frac{d}{dr}f(r)$.  In the Newtonian limit ($P_\parallel,P_\perp\ll\ed$, $G\,M\ll r$), $r$ is the radial distance from the center, $M(r)$ is the mass within radius $r$ and $\Phi(r)$ is the gravitational potential.  As in Newtonian gravity, the zero point of $\Phi$ is unspecified and can be chosen for convenience.  Regular stellar solutions have $r=0$ at the center of the star, a finite $M_\star\equiv M(r\rightarrow\infty)$ and require $2\,G\,M(r)<r$ so $M(0)=0$.  The metric asymptotes to the Schwarzschild form with mass $M_\star$ at large $r$. $M(r)$ can be interpreted as the gravitational mass inside radius $r$ and $M_\star$ the total gravitational mass of the star.

SM stars have very nearly isotropic pressure i.e. $|P_\parallel-P_\perp|\ll P\equiv\frac{1}{3}\,(P_\parallel+2\,P_\perp)$ in which case the
Tolman-Oppenheimer-Volkoff equation (TOV) equation\textcolor{red}{~\cite{PhysRev.55.364,PhysRev.55.374}} holds
\beq
\label{eq:TOVequation}
M'(r)=4\pi\,r^2\,\ed(r) \qquad
\frac{P'(r)}{\ed(r)+P(r)}=-\frac{G\,(M(r)+4\,\pi\,r^3\,P(r))}{r^2\,\left(1-\frac{2\,G\,M(r)}{r}\right)} .
\eeq
Cold SM stars have no atmospheres and the surface of the star at radius $r=R_\star$ is defined by $P(R_\star)=0$. The EoS defines a one parameter class of stellar solutions.  BSM stars may differ from SM stars not only in the EoS but they may have anisotropic pressure, $P_\parallel\ne P_\perp$.

\subsection{Balance of Chemical and Gravitational Forces}
\label{sec:ChemicalGravitationalBalance}

For any perfect fluid EoS, $P(\ed)$, one can define a conserved ``charge'' density: $n\propto\mathrm{exp}\int\frac{d\ed}{\ed+P(\ed)}$ which defines as conserved current $(n\,u^\mu)_{;\mu}=0$ where $u^\mu$ is the 4-velocity of the fluid.  This conservation law derives from perfect fluid energy momentum conservation: $\left((\ed+P)\,u^\mu u^\nu-P\,g^{\mu\nu}\right)_{;\nu}=0$, which also gives $\frac{P'(r)}{\ed(r)+P(r)}=-\Phi'(r)$ of eq.~\ref{eq:EinsteinEquations}.  If one re-expresses $\ed$ and $P$ in terms of $n$, i.e. $\ed(n)$ and $P(n)$, then one can define the {\it chemical potential} of this charge by
\beq
\cp(n)\equiv\ed'(n)=\frac{\ed(n)+P(n)}{n}
\label{eq:ChemicalPotential}
\eeq
where the last equality, the {\it adiabaticity equation}, follows from the definition of $n$.  One sees that $\cp'(n)=\frac{P'(n)}{n}$ and $\frac{P'(n)}{\ed(n)+P(n)}=\frac{\mu'(n)}{\mu(n)}$.  Thus  $\Phi'(r)=-\frac{P'(r)}{\ed(r)+P(r)}=-\frac{\cp'(r)}{\cp(r)} $ which has solution
\beq
\cp(r)\,e^{\Phi(r)}=\mathrm{constant}\ .
\label{eq:ConstantPotential}
\eeq
For hydrostatic equilibrium a spatial decrease in the gravitational potential, $\Phi$, must be compensated by a spatial increase in chemical potential, $\cp$.  This may be thought of as a balance of chemical and gravitational forces on the conserved charges.  This balance derives from energy momentum conservation, is true of any hydrostatic configuration whether spherical or not, and is more general than GR since it doesn't derive from Einstein's equations.  We will make use of this balance of forces below where the conserved charge is baryon number.

The conserved quantity $n$ can be interpreted as the number density of some ``chemical'' component of the fluid.  The chemical balance of eq.~\ref{eq:ConstantPotential} derives from pressure balance of eq.s~\ref{eq:EinsteinEquations}~\&~\ref{eq:TOVequation} so chemical and pressure balance encode the same information and are not separate principles.  This equivalence does require adiabaticity, i.e. that there is no energy leakage as one compresses fluid elements to higher densities, $\ed$.

\section{Ideal Gas + Yukawa (IGY) Equation of State}
\label{sec:IGY}

\subsection{Free Fermions}
\label{sec:freefermion}

In this section, we consider an ideal gas scenario where SM sector contains free fermions only.  For a free fermion species $f$ with (vacuum) mass $m_f$ and Yukawa coupling $g_f$ to $\phi$, the Dirac Lagrangian is given by
\beq
\mathcal{L}_\mathrm{eff}\supset  i\,\bar{\psi}_f\,\gamma^\mu\,\partial_\mu\,\psi_f-(m_f-g_f\,\tilde{\phi})\,\bar{\psi}_f\,\psi_f\ .
\label{eq:singleFermionLag}
\eeq
One sees that in a region with constant $\tilde{\phi}\ne0$  the fermion mass term differs from its vacuum value, i.e. $m_f\rightarrow m_f-g_f\,\tilde{\phi}$.  The dispersion relation for fermion $f$ is $\epsilon_f^2={\mkin_f}^2+|\mathbf{p}_f|^2$ where $\epsilon_f$ and $\mathbf{p}_f$ are the particle energy and momentum and 
\beq
\mkin_f(\tilde{\phi})\equiv |m_f-g_f\,\tilde{\phi}|
\label{eq:kinematicMass}
\eeq
is the \emph{kinematic mass} which we have defined to be non-negative.  Since the sign of the mass term in the Dirac Lagrangian is largely conventional the change of sign of $m_f-g_f\,\tilde{\phi}$ in eq.~\ref{eq:singleFermionLag}  for $\tilde{\phi}>m_f/g_f$  is not worrying.  The $\tilde{\phi}$ dependence of the kinematic mass of the SM fermion has important effects on EoS of the SM fermions. 

As shown in appendix~\ref{app:psibarpsi} for an isotropic particle distribution function,
\begin{eqnarray}
\label{eq:psibarpsiA}
\langle\bar{\psi}_f\psi_f\rangle
&=& n_f\,\langle\gamma^{-1}\rangle_f
=\frac{\ed_f-3\,P_f}{\mkin_f(\tilde{\phi})} \nonumber \\
\ed_f&=&\mkin_f(\tilde{\phi})\,n_f\,\langle\gamma\rangle_f \\
             P_{f}&=&\mkin_f(\tilde{\phi})\,n_f\,\langle\frac{\gamma^2-1}{3\,\gamma}\rangle_f \nonumber
\end{eqnarray}
where
$n_f$ is the fermion's number density, $\gamma$ is the Lorentz factor and $\langle\cdots\rangle_f$ gives the particle weighted average over the distribution function. $\ed_f$ and $P_f$ are the density and (isotropic) pressure of these particles including the $g_f\,\langle\bar{\psi}_f\psi_f\rangle\,\tilde{\phi}$ interaction term.  $n_f,\langle\gamma^{-1}\rangle_f$,  $\ed_{f}$ and $P_{f}$ are all frame dependent but their combination giving $\langle\bar{\psi}_f\psi_f\rangle$ is not.   There is no singularity in $\langle\bar{\psi}_f\psi_f\rangle$ when $\mkin_f\rightarrow0$ since in the massless limit $P_f\rightarrow\frac{1}{3}\,\ed_f$. 

Comparing eq.~\ref{eq:singleFermionLag} with eq.~\ref{eq:action}, the quantity $\tilde{n}$ given rise by the free fermions reads
\beq\label{eq:ntildeFermions}
\tilde{n}=\sum_f\,\frac{g_f}{\tilde{g}}\,n_{f}\langle\gamma^{-1}\rangle_{f}=
              \sum_f\,\frac{g_f}{\tilde{g}}\,\frac{\ed_{f}-3\,P_{f}}{\mkin_f(\tilde{\phi})}\ .
\eeq
In a GR context, $\tilde{n}$'s dependence on $\ed_{f}-3\,P_{f}$ is very convenient, as the gravitational field also couples to a fermion through $\ed_{f}$ and $P_{f}$ via their contribution to $T_{\mu\nu}$.  Eq.~\ref{eq:StressEnergy} now becomes
\beq
T_{\mu\nu}=T^{\tilde{\phi}}_{\mu\nu}+\sum_f \left(\left(\ed_{f}+P_{f}\right)\,u^{f}_\mu\,u^{f}_\nu
-P_{f}\,\mathrm{g}_{\mu\nu}+\tilde{P}^{f}_{\mu\nu}\right),
\label{eq:StressEnergyFermions}
\eeq
where $u^{f}_\mu$ is the center of momentum 4-velocity and $\tilde{P}^{f}_{\mu\nu}$ gives the anisotropic pressure tensor which is only non-zero if the momentum distribution is anisotropic in the $u^{f}_\mu$ frame.  The $\frac{\delta\tilde{n}}{\delta \mathrm{g}^{\mu\nu}}$ and $\tilde{g}\,\tilde{\phi}\,\tilde{n}\,\mathrm{g}_{\mu\nu}$ terms in eq.~\ref{eq:StressEnergy} is included in eq.~\ref{eq:StressEnergyFermions}, when one computes $\ed_{f}$ and $P_{f}$ with the kinematic mass $\mkin_f$ instead of $m_f$.  

In static stars, $u^{f}_\mu$ will be the stellar rest frame and we expect from SM interactions a nearly isotropic momentum distribution in this frame so  $\tilde{P}^{f}_{\mu\nu}\simeq0$ and
\beq
\ed=\ed_{\tilde{\phi}}+\sum_{f}\ed_{f}\ \qquad 
P               =               P_{\tilde{\phi}}+\sum_{f}               P_{f} \qquad 
P_\parallel-P_\perp=P^{\tilde{\phi}}_\parallel-P^{\tilde{\phi}}_\perp \ .
\label{eq:DensityPressureA}
\eeq
Only the scalar field contributes to the anisotropic pressure.

\subsection{Scalar Fields}
\label{sec:ScalarFields}

The $\phi$-sector Lagrangian including the interaction term is give by
\beq\label{eq:L_phi}
\mathcal{L}_\mathrm{eff}\supset \frac{\tilde{\phi}^{;\mu}\,\tilde{\phi}_{;\mu}}{2}-V(\tilde{\phi})+\tilde{g}\,\tilde{\phi}\,\tilde{n}~.
\eeq
The classical equation of motion (EoM) and stress-energy is given by
\beq
{\tilde{\phi}^{;\mu}}\phantom{}_\mu =-V'(\tilde{\phi})+\tilde{g}\,\tilde{n} \qquad
T^\phi_{\mu\nu}=\tilde{\phi}_{;\mu}\,\tilde{\phi}_{;\nu}+\mathrm{g}_{\mu\nu}\left(V(\tilde{\phi})-\frac{\tilde{\phi}^{;\xi}\,\tilde{\phi}_{;\xi}}{2}\right)~.
\label{eq:phiSector}
\eeq
For $\langle\phi\rangle=0$ to be a stable classical solution in vacuum ($\tilde{n}=0$) and contribute negligibly to the cosmological constant it is required that $V'(0)=0$, $V''(0)>0$ and $V(0)\simeq0$.

In the background of the static spherical star metric of eq.~\ref{eq:metricStressEnergy}, $\tilde{\phi}$ will have the same symmetries as the spacetime, i.e. with no temporal or azimuthal dependence, or $\tilde{\phi}(r)$.  This gives stress-energy contribution
\begin{eqnarray}
\ed_{\tilde{\phi}}(r)&=&+\frac{1-\frac{2\,G\,M(r)}{r}}{2}\,\tilde{\phi}'(r)^2+V(\tilde{\phi}(r)) \nonumber \\
               P_{\tilde{\phi}}(r)&=&-\frac{1-\frac{2\,G\,M(r)}{r}}{6}\,\tilde{\phi}'(r)^2-V(\tilde{\phi}(r))  \\
P^{\tilde{\phi}}_\parallel(r)-P^{\tilde{\phi}}_\perp(r)&=&\left(1-\frac{2\,G\,M(r)}{r}\right)\,\tilde{\phi}'(r)^2 \nonumber 
\label{eq:PhiEoS}
\end{eqnarray}
including the anisotropic pressure. The wave equation of  eq.~\ref{eq:phiSector} becomes
\beq
\tilde{\phi}''(r)=\frac{1}{2}\,\frac{\partial\,\mathrm{ln}\frac{e^{-2\Phi(r)}}{r^4\,\left(1-\frac{2\,G\,M(r)}{r}\right)}}{\partial r}\,\tilde{\phi}'(r)
+ \frac{V'(\tilde{\phi}(r))-\tilde{g}\,\tilde{n}(r)}{1-\frac{2\,G\,M(r)}{r}}
\label{eq:phiEquation}
\eeq 
For fixed $M(r)$ and $\tilde{n}(r)$ eq.~\ref{eq:phiEquation} is formally an inhomogeneous ODE for $\tilde{\phi}(r)$.  This is only formal since $\tilde{\phi}(r)$ may also be reflected in $M(r)$ and $\tilde{n}(r)$. If $V(\phi)=\frac{1}{2}\,m_\phi^2\,\phi^2$ then $\tilde{m}_\phi(\tilde{\phi})=m_\phi$ is constant and since $V'(\tilde{\phi}(r))=m_\phi^2\,\tilde{\phi}(r)$ this is a linear equation in $\tilde{\phi}$.  More generally eq.~\ref{eq:phiEquation}  is formally nonlinear in $\tilde{\phi}$.  For a given local value of $\tilde{\phi}$ we define the effective local $\phi$ mass by $\tilde{m}_\phi(r)\equiv V''(\tilde{\phi}(r))^{1/2}$.

As a convention we take $\phi=0$ to be  the vacuum value of $\phi$. We also impose various conditions on $V(\phi)$:
\beq
V(0)\approx0  \qquad V''(\tilde{\phi})>0 \qquad \left|\frac{V'' (\tilde{\phi})\,V (\tilde{\phi})}{V'(\tilde{\phi})^2}\right|\ngg1 \qquad
                                                                          \left|\frac{V'''(\tilde{\phi})\,V'(\tilde{\phi})}{V''(\tilde{\phi})^2}\right|\ngg1 ~.
\label{eq:Vassumptions}
\eeq 
which need only apply to the values of $\tilde{\phi}$ attained inside the star under study.  The 1st condition is assured by noting the observed cosmological constant is negligible compared to the density of stars. The 2nd condition avoids instabilities for large $\tilde{\phi}$ field values.  The other assumptions can be characterized as requiring that $V(\phi)$ not be ``unnaturally flat'' and are required from some of the conclusions we obtain below.

\subsection{Polarization}
\label{sec:polarization}

For the simplest potential, $V(\phi)=\frac{1}{2}\,m_\phi^2\,\phi^2$, in Minkowski space the classical $\phi$ EoM is the inhomogeneous Klein-Gordon equation
\beq
\ddot{\phi}-\nabla^2\phi+m_\phi^2\,\phi=\tilde{g}\,\tilde{n}
\label{eq:ScalarMinkowski}
\eeq
whose classical and quantum phenomenology are well known.  For a fixed (classical) static $\tilde{n}({\bf x})$, the lowest energy (ground state) solution is a Yukawa potential convolved with $\tilde{n}$:
\beq
\phi({\bf x},t)=\tilde{\phi}({\bf x})\equiv\frac{\tilde{g}}{4\pi}\int d^3{\bf x'}\,\frac{e^{-m_\phi\,|{\bf x}-{\bf x'}|}}{|{\bf x}-{\bf x'}|}\,\tilde{n}({\bf x'})\ .
\label{eq:YukawaGreen}
\eeq 
We call $\tilde{\phi}$ the ``polarization'' of $\phi$ by $\tilde{n}$.  To this one can add excitations 
$\propto e^{i\,({\bf k\cdot x}\pm\sqrt{|{\bf k}|^2+m_\phi^2}\,t)}$ which can be quantized into $\phi$ particles. Even with no such particle excitations there is a non-zero polarization: $\tilde{\phi}\ne0$.  Quantum and classical fluctuations in $\tilde{n}$ should be small when the number density of particles contributing to the integral is large, roughly $\tilde{n}\gg m_\phi^3$ for non- or mildly relativistic fermions.  

A polarization, $\tilde{\phi}$, tied to a static $\tilde{n}$ will exist in any static curved spacetime though the Green functions are more complicated.  If $\tilde{n}$ varies with time then $\tilde{\phi}$ will readjust itself to the new ground state on a timescale $\sim m_\phi^{-1}$.  In a star if the timescale of change is much slower than $m_\phi^{-1}$, then any excess energy above the ground state will go into coherent $\phi$ and SM matter oscillations.  Such excess energy above a ground state would eventually be radiated away by SM cooling processes until a static configuration is reached. If the temporal change has spectral content with angular frequency $|\omega|>m_\phi$ then coherent $\phi$ radiation may be emitted.  Since there is no conserved charge associated with $\tilde{n}$ this can include $\phi$ monopole radiation.
 
\subsection{Heavy, Light and Ultralight Scalars}
\label{sec:HeavyLightUltralight}

In flat space with $V(\phi)=\frac{1}{2}\,m_\phi^2\,\phi^2$ we see that in eq.~\ref{eq:YukawaGreen} if $\tilde{n}$ does not vary significantly over the size of the kernel, $m_\phi^{-1}$, then one can take $\tilde{n}$ out of the integral obtaining the approximation
\beq
\tilde{\phi}\approx\frac{\tilde{g}\,\tilde{n}}{m_\phi^2}\ .
\label{eq:LightScalarApproximation}
\eeq
This we refer to as the \textit{light scalar approximation} or LSA. The LSA corresponds to the case where the spatial gradients of $\tilde{\phi}$ contribute negligibly to the scalar wave equation, eq.~\ref{eq:phiEquation}, i.e.
\beq
V'(\tilde{\phi})&\approx\tilde{g}\,\tilde{n}
\label{eq:LightScalarApproximationGeneralized}
\eeq
which defines the LSA more generally when $V(\phi)\ne\frac{1}{2}\,m_\phi^2\,\phi^2$.  One might need to solve eq.~\ref{eq:LightScalarApproximationGeneralized} for $\tilde{\phi}$ numerically.  Spatial gradients add to the energy density and the ground state polarized field will minimize the gradients as much as possible. The gradients become negligible when both $\tilde{n}$ and $\tilde{m}_\phi$ not vary significantly over the range of the $\phi$-field, $\tilde{m}_\phi(\tilde{\phi})^{-1}=V_\phi(\tilde{\phi})^{-1/2}$.  In particular $|{\bf \nabla}\,\mathrm{ln}\,\tilde{n}|\ll\tilde{m}_\phi$ and  $|{\bf \nabla}\,\mathrm{ln}\,\tilde{m}_\phi|\ll\tilde{m}_\phi$. Using $\frac{V''(\tilde{\phi})}{V'(\tilde{\phi})}\nabla\tilde{\phi}\approx\nabla\ln\tilde{n}$ derived from eq.~\ref{eq:LightScalarApproximationGeneralized} these conditions become
\beq
|{\bf \nabla}\,\mathrm{ln}\,\tilde{n}|\ll V''(\tilde{\phi})^{1/2} \qquad 
\left|\frac{V'''(\tilde{\phi})\,V'(\tilde{\phi})}{V''(\tilde{\phi})^{5/2}}\,{\bf \nabla}\,\mathrm{ln}\,\tilde{n}\right|\ll1\ .
\eeq
Given the 1st condition the 2nd will be satisfied so long as the naturalness assumptions of eq.~\ref{eq:Vassumptions} is satisfied valid.  Eq.~\ref{eq:LightScalarApproximation} also applies to curved spacetime if the spacetime curvature scale is much smaller than ${\tilde{m}_\phi}^{-1}$  (roughly $G\,\ed\ll \tilde{m}_\phi^2$). 

No matter the form of $V(\phi)$ the LSA is only valid where the mean field approximation is valid so the SM matter distribution must be well approximated as a continuous field, or roughly $\tilde{m}^3_\phi(\tilde{\phi}(R))\ll n(R)$, where $n$ is the number density of the relevant fermions. The constraint gives the upper mass range of the LSA.  Scalars more massive than this we call {\it heavy scalars}.  Incoherent interactions of $\phi$ with SM particles will dominate for heavy scalars in contrast to lighter scalars where the interactions are coherent.  We do not consider heavy scalars in this paper.

\textit{Ultra-light scalars} have masses smaller than the LSA range.  In this case the relation of $\tilde{\phi}(R)$ to $\tilde{n}(R)$ is less trivial and generally must be determined numerically.  If $m(0)\,R_\star\lesssim1$ then the $\tilde{\phi}\ne0$ will extend well beyond the star contributing to a finite range 5th force which may effect orbiting objects.  Light scalars and heavy scalars in contrast to ultra-light scalars have their $\tilde{\phi}$ fields confined to the star and do not manifest long range forces.

For example in neutron stars the density varies on a scale of  $\sim 1\,$km with nuclear densities in the center and the density of normal liquids or solids ($n\sim\AA^{-3}$) at the surface.  Thus for the LSA to be valid throughout the star it is required that 
\beq
10^{-10}\,\mathrm{eV}\ll m_\phi\ll10^3\,\mathrm{eV}~.
\label{eq:lightScalarMassRangeNS}
\eeq
White dwarfs has similar surface densities but  $\sim10^3\times$ larger in size so the LSA is valid for
\beq
10^{-13}\,\mathrm{eV}\ll m_\phi\ll10^3\,\mathrm{eV}~.
\label{eq:lightScalarMassRangeWD}
\eeq
The partition of scalar masses into heavy, light and ultra-light  regimes is context dependent.  The LSA may be valid in the core of a star but not near its surface.

\subsection{Light Scalar Equation of State}

The $\phi$ stress-energy given by eq.~\ref{eq:PhiEoS} depends on two terms $|\nabla\tilde{\phi}|^2$ and $V(\tilde{\phi})$.  In the LSA, using eq.~\ref{eq:LightScalarApproximationGeneralized}, the ratio of these two terms is 
\beq
\frac{|\nabla\tilde{\phi}|^2}{|V(\tilde{\phi})|}=\left|\frac{V'(\tilde{\phi})^2}{V''(\tilde{\phi})\,V(\tilde{\phi})}\,
\frac{|\nabla\,\mathrm{ln}\,\tilde{n}|^2}{V''(\tilde{\phi})}\right|\ll1~,
\label{eq:Vdomination}
\eeq
where we have used the LSA requirement $|\nabla\,\mathrm{ln}\,\tilde{n}|^2\ll |V''(\tilde{\phi})|$ and a naturalness assumption of eq.~\ref{eq:Vassumptions}.  Thus the LSA implies
\beq
\label{eq:PhiStressEnergyLSA}
\ed_{\tilde{\phi}}\approx-P_{\tilde{\phi}}\approx V(\tilde{\phi}) \qquad P^{\tilde{\phi}}_\parallel-P^{\tilde{\phi}}_\perp\approx 0
\eeq
so even the scalar stress energy is nearly isotropic and one may use the TOV equation~\ref{eq:TOVequation} to compute stellar structure.

\subsubsection{IGY Equation of State}
\label{sec:IGYEoS}

To understand the qualitative aspects of how a linearly coupled $\phi$ effects the EoS of matter consider a cold ideal gas of a single fermion, $f$, Yukawa coupled to a scalar with the simplest scalar potential: $V(\phi)=\frac{1}{2}\,m_\phi^2\,\phi^2$.  This EoS we denote IGY for ideal gas + Yukawa.  Take $\tilde{g}=g_f$ so $\tilde{n}=\tilde{n}_f$. Given eq.s~\ref{eq:psibarpsiA}, \ref{eq:ntildeFermions}~\&~\ref{eq:LightScalarApproximation} we find for an arbitrary distribution function
\begin{eqnarray}
\label{eq:IGY_EandP}
\ed&=+V(\tilde{\phi})+\ed_f
                   =&+\frac{1}{2}\,\left(\frac{g_f\,n_f\,\langle\gamma^{-1}\rangle_f}{m_\phi}\right)^2
                   +\left(m_f-\frac{g_f^2\,n_f}{m_\phi^2}\,\langle\gamma^{-1}\rangle_f\right)
                   \,n_f\,\langle\gamma\rangle_f \nonumber \\
              P&=-V(\tilde{\phi})+P_f
                   =&-\frac{1}{2}\,\left(\frac{g_f\,n_f\,\langle\gamma^{-1}\rangle_f}{m_\phi}\right)^2
                   +\left(m_f-\frac{g_f^2\,n_f}{m_\phi^2}\,\langle\gamma^{-1}\rangle_f\right)
                   \,n_f\,\langle\frac{\gamma^2-1}{3\,\gamma}\rangle_f       \ .
\end{eqnarray}
Specializing to an ideal cold degenerate Fermi gas using the dimensionless functions, $\xi$ and $\varphi$, defined in eq.s~\ref{eq:E}~\&~\ref{eq:P} we obtain
\beq
\label{eq:n_free}
\xi(x)                  =\frac{x^2}{3\pi^2}\,\langle\gamma\rangle_f  \qquad
\varphi(x)           =\frac{x^2}{3\pi^2}\,\langle\frac{\gamma^2-1}{3\,\gamma}\rangle_f \qquad
\eta(x)               =\frac{x^2}{3\pi^2}\langle\gamma^{-1}\rangle_f=\xi(x)-3\,\varphi(x) \ .
\eeq
in terms of $x\equiv\beta_\mathrm{F}\gamma_\mathrm{F}=p_\mathrm{F}/\mkin_f$ and $n_f=\frac{x^2}{3\pi^2}\,{\mkin_f}^3$.  Here $p_\mathrm{F}$ is the Fermi momentum.  It is convenient to parameterize the EoS in terms of $x$  rather than $n_f$.

Combining eq.s~\ref{eq:kinematicMass},~\ref{eq:LightScalarApproximation}~\&\ref{eq:n_free} gives an expression for the kinematic mass in terms $x$:
\beq
\mr(x)\equiv\frac{\mkin_f}{m_f}=1-\geff_f^2\,\mr(x)^3\,\eta(x) \qquad 
\geff_f\equiv \frac{g_f\,m_f}{m_\phi}\ .
\eeq
There is a single positive real solution for $\mr$ to this cubic equation
\beq
\mr(x)\equiv\frac{\mkin_f}{m_f}
          =\frac{\left(\sqrt{1+\frac{4}{27\,\geff_f^2\,\eta(x)}}+1\right)^\frac{1}{3}
                   -\left(\sqrt{1+\frac{4}{27\,\geff_f^2\,\eta(x)}}-1\right)^\frac{1}{3}}
                                         {(2\,\geff_f^2\,\eta(x))^\frac{1}{3}}  \in [0,1) ~ .
\label{eq:muplus}
\eeq
The range of $\tilde{\phi}$ only spans $0\le\tilde{\phi}<\tilde{\phi}_\mathrm{max}^f\equiv m_f/g_f$ so $\mr(x)\ge0$.  In this model $g_f$ only enters the EoS through $\geff_f$, the {\it effective coupling}. 

Combining eq.s~\ref{eq:IGY_EandP},~\ref{eq:n_free}~\&~\ref{eq:muplus} the IGY EoS is given by
\begin{align}
\label{eq:EoS_IGY}
\frac{\ed_\mathrm{IGY}(x)}{m_f^4}&=\mr(x)^4\,\xi       (x)+\mathcal{V}_\mathrm{IGY}(x) &
\frac{n_f(x)             }{m_f^3}&=\mr(x)^3\,\frac{x^3}{3\,\pi^2}           &
 \mathcal{V}_\mathrm{IGY}(x)&=\frac{{\geff_f}^2}{2}\,\mr(x)^6\,\eta(x)^2 \nonumber \\
\frac{              P_\mathrm{IGY}(x)}{m_f^4}&=\mr(x)^4\,\varphi(x)-\mathcal{V}_\mathrm{IGY}(x) &
\frac{ \tilde{n}_f(x) }{m_f^3}&=\mr(x)^3\,\eta(x) &
 \frac{E_\mathrm{F}(x)}{m_f}&=\mr(x)\,\sqrt{1+x^2} \ .
\end{align}
This is simplified further in appendix \ref{app:IGYsimplified}. Here $\mathcal{V}_\mathrm{IGY}$ contains the $\tilde{\phi}$ contribution to the energy density $\ed$ and pressure $P$.  EoSs are usually expressed as $P(\ed)$ but there is no analytic expression for either of these in IGY and 
the IGY EoS is only defined implicitly by eq.s~\ref{eq:EoS_IGY}.  Also given is the fermion Fermi energy, $E_\mathrm{F}\equiv\sqrt{{\mkin_f}^2+p_\mathrm{F}^2}$.  One can verify {\it adiabaticity}
\beq
\cp_f(x)\equiv\frac{\partial\ed_\mathrm{IGY}}{\partial n_f}=\frac{{\ed_\mathrm{IGY}}'(x)}{{\tilde{n}_f}'(x)}
=\frac{\ed_\mathrm{IGY}(x)+P_\mathrm{IGY}(x)}{\tilde{n}_f(x)}
=E_\mathrm{F}(x)
\label{eq:IGYadiabaticity}
\eeq
which is the 1st law of thermodynamics (energy conservation) $dU=-P\,dV+T\,dS+\cp_f\,dN_f$ for adiabatic compression ($dS=0$) when the number of fermions is conserved ($dN_f=0)$.  Here $U$, $V$ and $N_f$ are the energy, volume and number of fermions for a fluid element and $\cp_f$ is the fermion chemical potential.  For IGY $\cp_f=E_\mathrm{F}$ since fermions are added to or subtracted from the Fermi surface.

The $\phi$ contribution to the the EoS in eq.s~\ref{eq:PhiStressEnergyLSA}~\&~\ref{eq:EoS_IGY} has the form of a cosmological constant which goes to zero at low densities, $n_f\rightarrow0$, and saturates,  
$\mathcal{V}_\mathrm{IGY}\rightarrow\frac{1}{2\,\geff_f^2}$, at high densities, $n_f\rightarrow\infty$.  Since the fermion density and pressure grow unbounded as  $n_f\rightarrow\infty$ the $\phi$ contribution becomes negligible at both high and low densities.  It is only at intermediate densities, as the fermion transitions from non-relativistic to ultra-relativistic behavior, that the Yukawa coupling will effect the EoS.  The large decrease in the fermion kinematic mass at very high density has no effect since the fermion mass is insignificant for ultra-relativistic particles. The fermions do become relativistic at lower densities the larger the value of $\geff_f$ is due to the more rapid decrease in $\tilde{m}_f$ with density.  This combined with the negative scalar field pressure which comes to dominate at intermediate densities results in a large $\geff_f$ dependence of the EoS for the range of intermediate densities.  The asymptotic behavior of the IGY EoS is independent of $\geff_f$:
\beq
\ed_\mathrm{IGY}\rightarrow
\begin{cases}
m_f\,n_f                                            & \tilde{n}_f\rightarrow 0 \\
\left(\frac{9\pi}{8}\right)^\frac{2}{3}\,{n_f}^\frac{4}{3} & \tilde{n}_f\rightarrow \infty
\end{cases}
\qquad
P_\mathrm{IGY}\rightarrow
\begin{cases}
\frac{1}{5}\,(3\,\pi^2)^\frac{2}{3}\,\frac{{n_f}^\frac{5}{3}}{m_f}        & \tilde{n}_f\rightarrow 0 \\
\frac{1}{3}\,\left(\frac{9\pi}{8}\right)^\frac{2}{3}\,{n_f}^\frac{4}{3} & \tilde{n}_f\rightarrow \infty
\end{cases} \ ,
\eeq
the same as a degenerate Fermi gas with zero Yukawa coupling.  At high density the fermion kinematic mass becomes small but does not go negative since $\tilde{\phi}>\frac{m_f}{g_f}$.

\begin{figure}
\centering
\includegraphics[width=\textwidth]{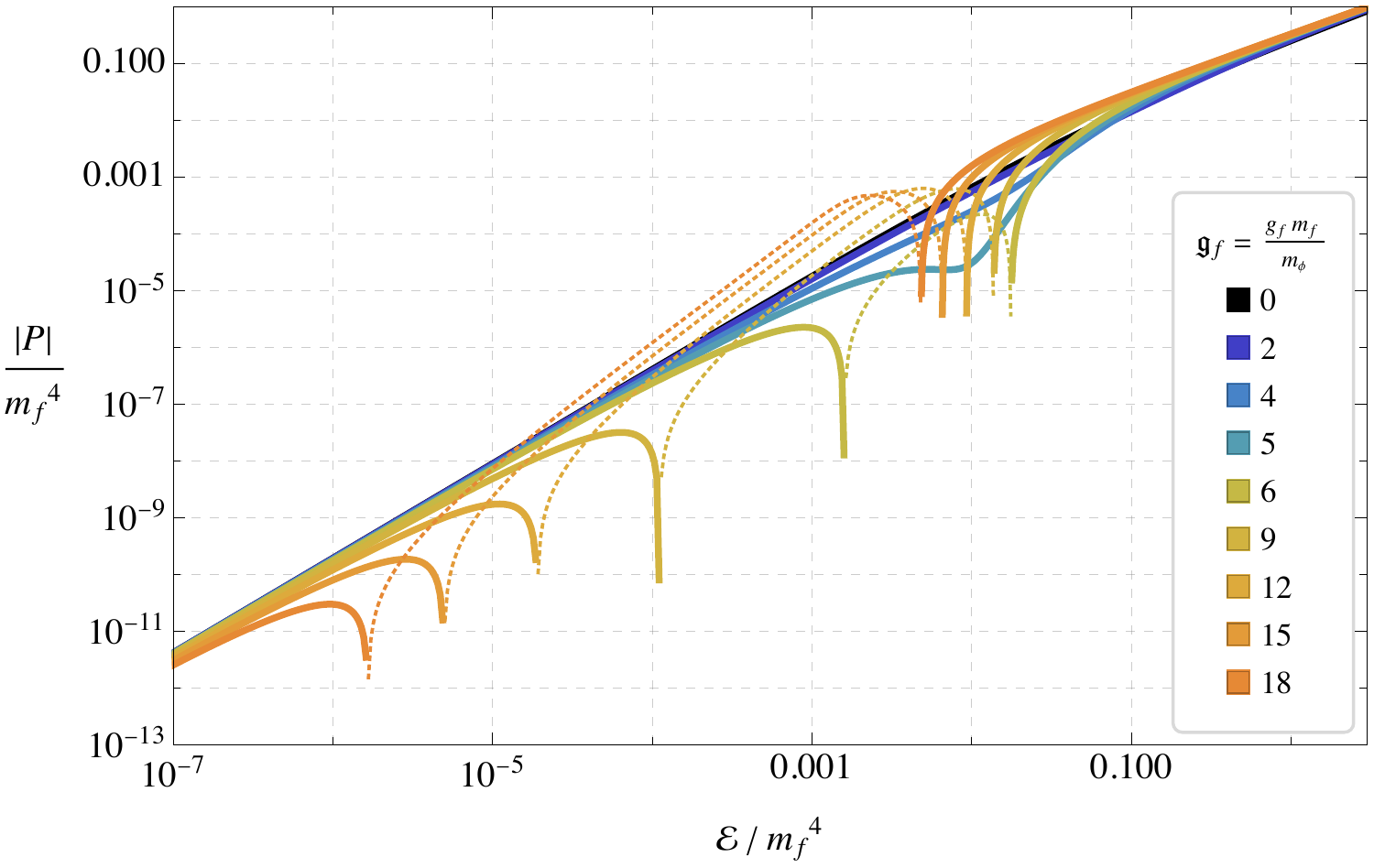}
\caption{Plotted are the ideal gas + Yukawa (IGY) equation-of-state (EoS), $P(\ed)$, from eq.~\ref{eq:EoS_IGY} for various values of the coupling $\geff_f\equiv\frac{g_f\,m_f}{m_\phi}$ as indicated in the legend.  Negative pressures are plotted as dashed curves.  The partially obscured black curve is the idea gas EoS with no Yukawa coupling.}
 \label{fig:EoSIGY1}
\end{figure}

The IGY EoS, $P(\ed)$, is plotted for various values of $\geff_f$ in fig.~ \ref{fig:EoSIGY1}.  A significant suppression of pressure relative to the $\geff_f=0$ EoS occurs for $\geff_f\gtrsim3$ and only for a range of $\ed$ as illustrated. Pressure suppression is more pronounced and grows more rapidly for larger $\geff_f$ leading to larger suppression at lower densities and negative pressures for $\geff_f\gtrsim5$ .  Pressure suppression is the main consequence of a Yukawa coupling and is caused by the combination of decreasing $\tilde{m}_f$ and the negative contribution of the scalar field polarization $\tilde{\phi}$. 

\section{Equation of State Anomalies}
\label{sec:anomalies}

The relativistic fluid described by the IGY EoS of eq.~\ref{eq:EoS_IGY} has many conventional properties 
\begin{itemize}
\item energy density increases with particle density: $\frac{\partial\ed}{\partial n_f}>0$,
\item no anomalously large sound speeds: $\frac{\partial P}{\partial\ed}<\frac{1}{3}$,
\item weak energy condition: $\ed\ge0$,
\item null energy condition: $\ed+P\ge0$,
\item dominant energy condition: $\ed\ge|P|$ and
\item adiabaticity: $\frac{\partial\ed}{\partial n_f}=\frac{\ed+P}{n_f}$.
\end{itemize}
Furthermore as discussed previously the small and large density behavior are precisely those of an ideal cold Fermi gas of a single fermion species as indicated above.

In addition to conventional behavior the IGY EoS exhibits {\it anomalous} behavior:
\begin{enumerate}
\item $\frac{\partial P}{\partial\ed}<0$ for $\geff_f>\geff_\mathrm{crit}=4.987\ldots$ and $\ed^-_\sim(\geff_f)\le\ed\le\ed^+_\sim(\geff_f)$,
\item $P<0$ for $\geff_f>\geff_0=5.124\ldots$ and
$\ed^-_0(\geff_f)\le\ed\le\ed^-_0(\geff_f)$,
\item $\ed+3P<0$ for $\geff_f>28.094\ldots$ gravitational repulsion - a violation of the strong energy condition.
\end{enumerate}
each of which persist only over a finite range of densities for a given $\geff_f$.   While these anomalous states of matter do exist for IGY (some hydrodynamically unstable for wavenumbers $k\lesssim\,m_\phi$) it is argued below that they do not actually occur in a stellar environment.   While this pressure behavior is anomalous for stellar matter it is not uncommon in condensed matter physics.  As with condensed matter qualitatively new phenomena can arise because of these anomalies and can be understood in terms of different phases of matter as is now described.

\subsection{Forbidden, Soft and Hard Matter}
\label{sec:ForbiddenSoftHard}

As one increases $\geff_f$ first anomaly 1 occurs first, then 1 \& 2 and finally 1 \& 2 \& 3. The onset of these pressure anomalies occur at {\it critical point}: $\geff_f=\geff_\mathrm{crit}$ and $\ed_\mathrm{crit}=\ed^-_\sim(\geff_\mathrm{crit})=\ed^+_\sim(\geff_\mathrm{crit})=0.006197...$
when the fermions are non-relativistic and when the $\phi$ contributes only a small fraction of the total: $\ed_\phi/\ed\approx0.06$.  This is possible because the fermion pressure is so small, $P_f/\ed_f\approx0.07$ whereas $P_{\tilde{\phi}}/\ed_{\tilde{\phi}}=-1$ in the LSA.  The rapid onset of anomalous behavior with increasing $\geff_f$ means that stellar structure can be very sensitive to the value of this parameter. 

\begin{figure}
\centering
\includegraphics[width=\textwidth]{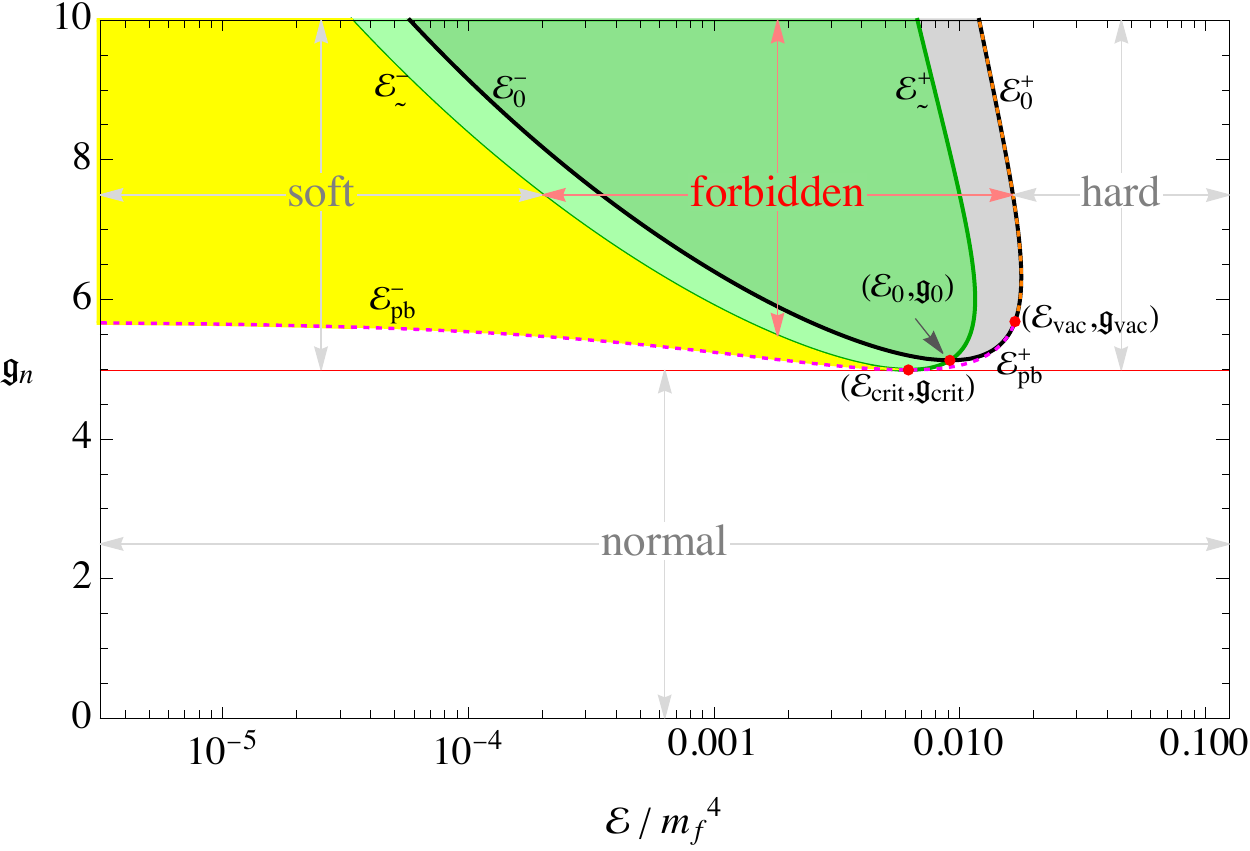}
\caption{
Shown is a ``phase diagram" for the IGY equation of state, $P(\ed)$, as a function  of the energy density $\ed$ in units of ${m_f}^4$ and the effective coupling $\geff_f\equiv\frac{g_f\,m_f}{m_\phi}$.  Since IGY is restricted to cold ($T=0$) matter the phase is a function of one parameter, $\ed$, but this dependence differs for different couplings $\geff_f$.   Shown are {\it normal}, {\it forbidden}, {\it soft} and {\it hard} phases as described in the text.  The forbidden phase is not to be found in stars either because of hydrodynamic instability, $\frac{\partial P}{\partial\ed}<0$ (green shaded) or negative pressure (gray shaded).  The onset of instability occurs at $(\ed_\mathrm{crit},\geff_\mathrm{crit})$ and negative pressure at $(\ed_0,\geff_0)$; these two anomalies are restricted to  densities $\ed_\sim^-<\ed< \ed_\sim^+$ and  $\ed_0^-<\ed< \ed_0^+$, respectively; with lower allowed densities referred to as {\it soft} and higher allowed densities as {\it hard}.  If $\geff_\mathrm{crit}<\geff_f<\geff_\mathrm{vac}$ hard and soft matter can form a stable boundary in pressure and chemical equilibrium at soft density $\ed^-_\mathrm{pb}$ and hard density $\ed^+_\mathrm{pb}$ (pink dashed curve).  If $\geff_f\ge\geff_\mathrm{vac}$ hard matter with density $\ed_0^+$ (orange dashed curve) can form a stable boundary in pressure equilibrium with vacuum.  Soft matter in the shaded yellow region is metastable to a transition to hard matter.  If $\geff_f<\geff_\mathrm{crit}$ there is no anomalous behavior which we refer to as {\it normal}.
 }
\label{fig:EoSIGY2}
\end{figure}

For $\geff_f<\geff_\mathrm{crit}$ matter at all densities behave normally (no anomalies) which we refer to as {\it normal} matter.  For $\geff_f>\geff_\mathrm{crit}$ some points in the $\geff_f$-$\ed$phase plane cannot exist inside a star.  Such phases of matter we refer to as {\it forbidden} and are indicated in fig.~\ref{fig:EoSIGY2}.  In order to be hydrodynamically stable one requires $\frac{\partial P}{\partial\ed}\ge0$ which is not satisfied for $\geff_f>\geff_\mathrm{crit}$ in the density range $\ed^-_\sim(\geff_f)\le\ed\le\ed^+_\sim(\geff_f)$.  This instability region is forbidden. Since a star has zero pressure at its surface, as indicated by the TOV eq.~\ref{eq:TOVequation}, the gravitational overburden will cause the pressure to rise as one goes from the surface toward the center and it follows that $P\ge0$ in the star.  This is not satisfied for $\geff_f>\geff_0$ in the density range $\ed^-_0(\geff_f)\le\ed\le\ed^-_0(\geff_f)$.  The negative pressure region is also forbidden.  Since 
$\ed^-_\sim(\geff_f)<\ed^-_0(\geff_f)<\ed^+_\sim(\geff_f)<\ed^+_0(\geff_f)$
the instability region and the negative pressure region overlap forming the entire forbidden region.  These two regions in the $\geff_f$-$\ed$ plane are shown in  fig.~\ref{fig:EoSIGY2}.

For $\geff_f>\geff_\mathrm{crit}$ the forbidden region separates a low density phase of matter that we call {\it soft} from a higher density phase of matter that we call {\it hard}.  This terminology derives from the quantity $\frac{\partial\,\mathrm{ln}P}{\partial\,\mathrm{ln}\ed}$ which is a measure of hardness of an EoS.  The hard phase can have large values $\frac{\partial\,\mathrm{ln}P}{\partial\,\mathrm{ln}\ed}$ which becomes infinite at $\ed=\ed_0^+$ whereas soft matter has $0\le\frac{\partial\,\mathrm{ln}P}{\partial\,\mathrm{ln}\ed}\le\frac{5}{3}$.

If $\geff_f\le\geff_\mathrm{crit}$ then only a normal phase exists, all $\ed$ are possible since $\frac{\partial P}{\partial\ed}\ge0$ and $P>0$.  In this $\geff_f$-range a star can contain any $\ed$ allowed by gravitational stability.   This is not true if $\geff_f>\geff_\mathrm{crit}$ where a star can consist of either a  soft phase or a hard phase or both but there is a forbidden range of $\ed$.  Soft, forbidden and hard phases occur whenever $P(\ed)$ increases from zero at small $\ed$, decreases for intermediate $\ed$ and increases again for large $\ed$.  Below we describe how a hard and soft phase can coexist in hydrostatic equilibrium and how a hard phase can interface with vacuum in hydrostatic equilibrium with no intermediate soft phase.

The phases of Yukawa coupled cold matter is somewhat analogous to the behavior of liquid and gas phases for ordinary materials such as water:  the hard phase is analogous to a liquid phase and the soft phase analogous to a gas phase.  These two phases only exist in the coupling range $\geff_f>\geff_\mathrm{crit}$.  One can consider a hard phase as having certain (equilibrium) ``vapor pressure'' for the soft phase with which it is in equilibrium.  For a small range of couplings  $\geff_\mathrm{crit}<\geff_f<\geff_\mathrm{vac}$ this vapor pressure is non-zero allowing hard and soft phases to coexist ($\geff_\mathrm{vac}$ is the value of $\geff_\mathrm{n}$ where $\cp_f(\ed_0^+)=\ce_f(\ed_0^+)=m_f)$.  It is possible that the soft phase can exist in a metastable state with pressure larger than the equilibrium vapor pressure.  This is analogous to supersaturation of water vapor ($>100\%$ humidity).  The equilibrium vapor pressure is 0 if $\geff_f>\geff_\mathrm{vac}$ and all low density soft matter is metastable in this case.   There is a large barrier for the metastable soft phase to transition to lower energy stable hard phase.  One way to seed such a phase transition is to compress the soft phase enough to reach the instability region $\ed_f>\ed^-_\sim$ over a volume greater than $m_\phi^{-3}$. The instability will grow to produce some hard matter which can seed the transition of all the connected metastable soft matter to hard matter.

\subsection{Phase Boundaries}
\label{sec:Interfaces}

Phase boundaries allow matter of very different densities to exist alongside each other, allowing stars or other objects in hydrostatic equilibrium to avoid the forbidden unstable region.  One can understand the phase boundaries on macroscopic scales in terms of jump conditions between two allowed phases arrived at from general principles or one can derive these jump conditions from a microscopic analysis.  We begin with a microscopic description on the length scales of the force carrier, ${m_\phi}^{-1}$ abandoning the LSA which is only valid on scales $\gg{m_\phi}^{-1}$.

A phase boundary is a hydrostatic configuration with planar symmetry separating matter with two asymptotic bulk properties.  The transition layer is assumed to be narrow enough so that gravitational tidal forces from one side to the other are negligible.  In flat space stress-energy conservation, ${T^{\mu\nu}}_{;\nu}=0$, for hydrostatic solutions have constant pressure, ${\bf \nabla}P=0$.  In terms of the scalar and fermionic pressure from eq.s~\ref{eq:psibarpsiA},\,\ref{eq:phiSector}\,\&\,\ref{eq:IGY_EandP} the pressure profile is given by
\beq
\label{eq:}
P=P_f(z)+P^\parallel_\phi(z)=(m_f-g_f\,\tilde{\phi}(z))^4\,\varphi(x(z))+\frac{1}{2}\,\left(\tilde{\phi}'(z)^2-{m_\phi}^2\,\tilde{\phi}(z)^2\right)
\label{eq:constantPressure1D}
\eeq
where $z$ is the Euclidean coordinate orthogonal to the transition plane and ${}'\equiv\frac{d}{dz}$.  Using eq.~ \ref{eq:ScalarMinkowski} the equation of hydrostatic equilibrium becomes $P_f'(z)=\geff_f\,\tilde{n}_f(z)\,\tilde{\phi}'(z)$ which balances the scalar field gradient force by a pressure gradient in the fermion fluid. From this one can show (for an ideal cold fermion gas) that the the Fermi energy, $E_\mathrm{F}=(m_f-g_f\,\tilde{\phi}(z))\,\sqrt{1+x(z)^2}$, is constant, relating $x(z)$ to $\tilde{\phi}(z)$.  Choosing the $+z$ to be the direction in which $\tilde{\phi}$ increases one obtains the ODE describing planar hydrostatic equilibrium without gravity:
\beq
\label{eq:}
\tilde{\phi}'(z)=\sqrt{2\,P-2\,(m_f-g_f\,\tilde{\phi}(z))^4\,\varphi\left[\sqrt{\left(\frac{E_\mathrm{F}}{m_f-g_f\,\tilde{\phi}(z)}\right)^2-1}\right]
+{m_\phi}^2\,\tilde{\phi}(z)^2}
\label{eq:hydrostatic1D}
\eeq
whose solution is characterized by the two physical constants $P$ and $E_\mathrm{F}$ plus a third which translates the configuration in $z$.  Only when $P$, $E_\mathrm{F}$ and $\tilde{\phi}$ take on values of any IGY solution (eq.s~\ref{eq:EoS_IGY}~\&\,\ref{eq:LightScalarApproximation}) will $\tilde{\phi}'(z)=0$ so only IGY EoS's can describe any asymptotic ``bulk'' properties of cold matter. 

\begin{figure}
\centering
\includegraphics[width=\textwidth]{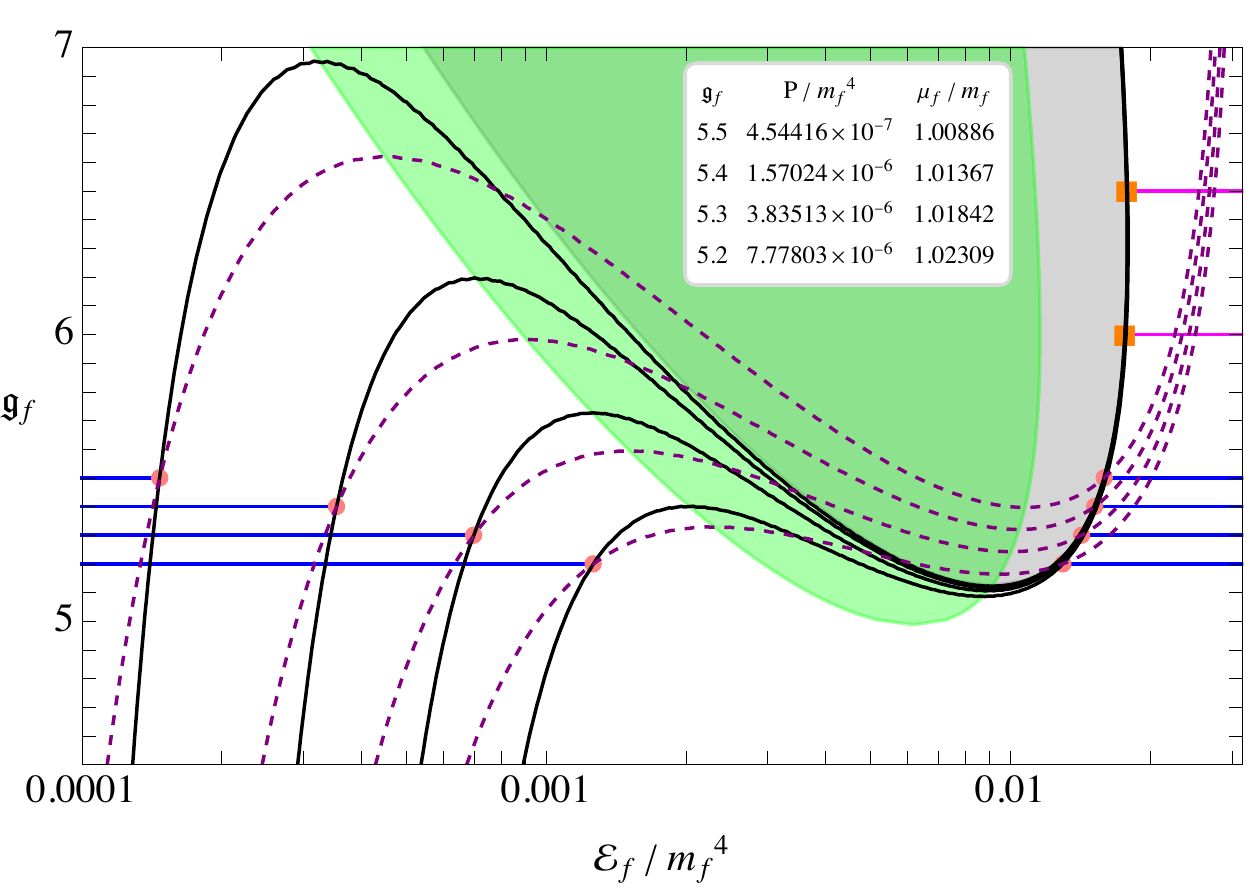}
\caption{
This graph illustrates the phase boundaries which are allowed by the IGY EoS.   Superposed on an expanded view of the phase diagram of fig.~\ref{fig:EoSIGY2} are iso-pressure ($P$ - black) and iso-chemical potential ($\cp_f$ - dashed purple) curves at values of these quantities which form stable interfaces between soft and hard matter at four values of the effective coupling $\geff_f=g_f m_f/m_\phi$.  The values of $\geff_f$, $P$ and $\cp_f$ are given in the inset.  The condition for a phase boundary is that the pressure and chemical potential do not change across the boundary for a fixed value of $\geff_f$.   One can see that this condition is satisfied for these values.   These phase boundaries allow the matter in a star to increase from low densities along the blue horizontal lines and then jump from the pink dot on the left to the pink dot on the right at much higher density from which the density can increase further.  This jump is from soft matter to hard matter circumventing the forbidden region.  Soft/hard phase boundaries do not exist when $\geff_f>\geff_\mathrm{vac}$ but in this case hard matter can form a stable interface with vacuum.  This vacuum/hard matter interface is indicated by the orange squares where the density jumps from zero to a larger (hard) value and further increases inside the star along the magenta lines.  The pink dots lie along the pink dashed $\ed_\mathrm{pb}^\pm$ curves in fig.~\ref{fig:EoSIGY2} while the orange squares lie along the orange dashed $\ed_0^+$ curve.
 }
\label{fig:InterfaceGraph}
\end{figure}

\subsubsection{Connecting Hard and Soft Matter}
\label{sec:SoftHardInterface}

A trivial solution to eq.~\ref{eq:hydrostatic1D} is any uniform IGY EoS where $P$, $E_\mathrm{F}$ and $\tilde{\phi}$ are given by the same IGY phase but this is not a boundary between two different phases.  Values of $P$ and $E_\mathrm{F}$ corresponding to phase boundaries must have {\it different} asymptotic bulk properties corresponding to two different phases, i.e. two different  $\tilde{\phi}$ values with the same $P$ and $E_\mathrm{F}$.  For an adiabatic EoS the chemical potential have extrema when $0=\frac{\partial\cp_f}{\partial n_f}=\frac{1}{n_f}\,\frac{\partial P}{\partial n_f}$ or $P'(\ed)=0$ (since $\frac{\partial\ed}{\partial n_f}>0$).  For normal matter ($\geff_f<\geff_\mathrm{crit}$)  $P'(\ed)>0$ and thus $\frac{\partial\cp_f}{\partial n_f}>0$ so no two phases have the same $P$ or the same $\cp_f$ and a phase boundary is not possible.  However when $\geff_f>\geff_\mathrm{crit}$  both $P$ and $\cp_f$ have a maximum and a minimum at $\ed=\ed_\sim^-$ and $\ed_\sim^+$, respectively so different phases can have the same $P$ and/or the same $E_\mathrm{F}$ and phase boundaries are possible. Mathematically, from eq.~\ref{eq:EoS_IGY}, phase boundaries are described by solutions of
\begin{eqnarray}
       -\frac{1}{2}\,\geff_f^2\,\mr(x_-)^6\,\eta(x_-)^2+\mr(x_-)^4\,\varphi(x_-)
&=&-\frac{1}{2}\,\geff_f^2\,\mr(x_+)^6\,\eta(x_+)^2+\mr(x_+)^4\,\varphi(x_+)   \nonumber \\
\mr(x_-)\,\sqrt{1+{x_-}^2}&=&\mr(x_+)\,\sqrt{1+{x_+}^2}
\label{eq:JumpConditions}
\end{eqnarray}
where $x_\pm\equiv x(\pm\infty)$ and $x_-<x_+$. The asymptotic energy densities of these solutions are
\beq
\ed^\pm_\mathrm{bp}(\geff_f)\equiv \ed(\pm\infty)\
= m_f^4\,\left(+\frac{1}{2}\,\geff_f^2\,\mr(x_\pm)^6\,\eta(x_\pm)^2+\mr(x_\pm)^4\,\xi(x_\pm)\right) 
\label{eq:EnergyJump}
\eeq
which can differ by orders of magnitude.  $\ed^\pm_\mathrm{bp}$ lie along the pink dashed curves in fig.~\ref{fig:EoSIGY2} which meet at the critical point.  There is a single solution of eq.s~\ref{eq:JumpConditions} only for $\geff_\mathrm{crit}<\geff_f\le\geff_\mathrm{vac}$.   Fig.~\ref{fig:InterfaceGraph} shows four examples of  such solutions for different values of $\geff_f$.  The width of the transition layer is several ${m_\phi}^{-1}$ which by assumption is much smaller than the length scale of the matter configuration under the LSA.  The asymptotic $x_-$ IGY phase is always soft matter and the asymptotic $x_+$ is always hard matter so these phase boundaries allow the fluid to jump over the forbidden region.   The phase boundaries occur at particular values of soft and hard matter densities which are $\geff_f$ dependent.  The pressure at these phase boundaries are always positive.  

One could have foregone this microscopic analysis since from general principles a stable phase boundary should obey the Gibbs jump conditions 1) pressure balance: the pressure of the two phases must be equal, for hydrostatic equilibrium and 2) chemical equilibrium: the chemical potential of the fermions (the Fermi energy in this case) must be equal  otherwise fermions would flow into the phase with lower $\cp_f$ causing the boundary to move.  This macroscopic analysis at the level of LSA only requires solutions of the jump conditions of eq.~\ref{eq:JumpConditions} and would treat the phase boundary as being discontinuous.

\begin{figure}
\centering
\includegraphics[width=\textwidth]{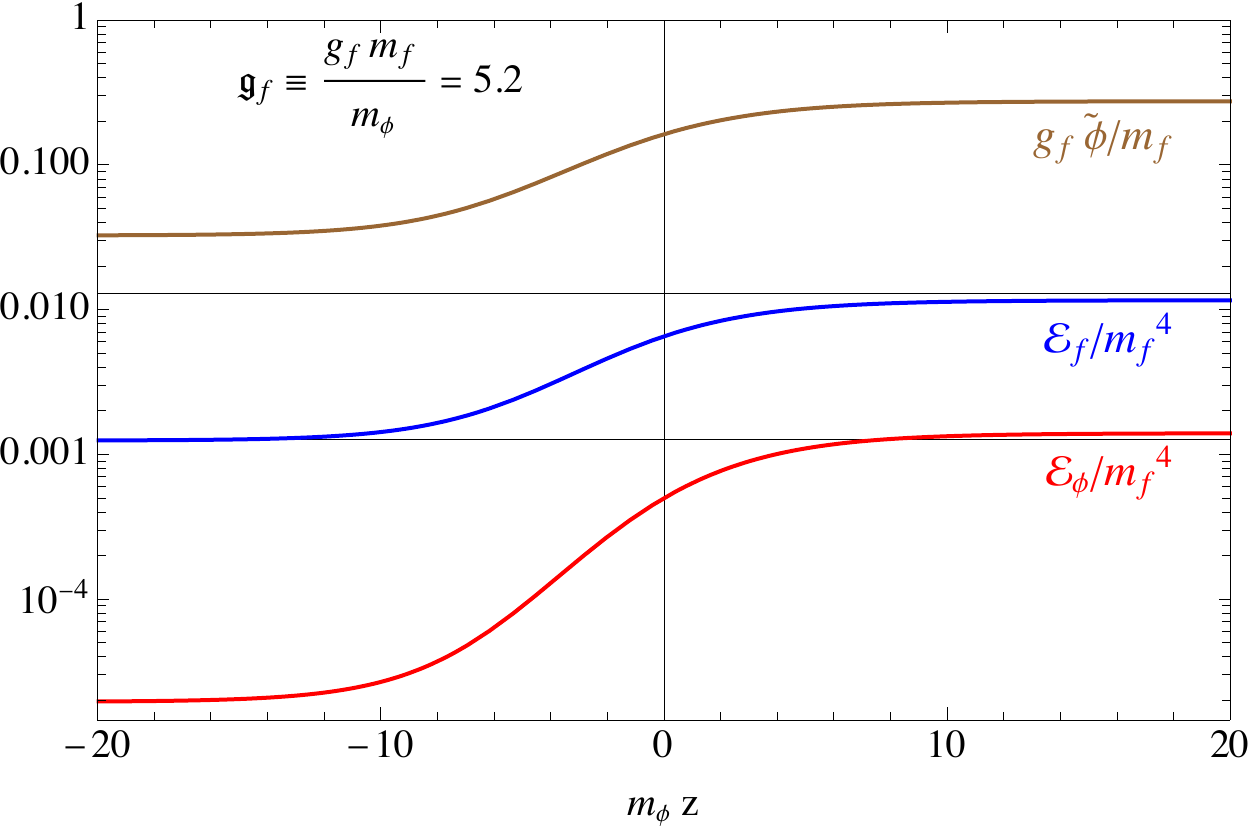}
 \caption{Plotted is the profile of a 1-D phase boundary between soft with hard matter for $\geff_f=5.2$ as described in \S\ref{sec:SoftHardInterface}.  A bulk IGY soft state is reached as $z\rightarrow-\infty$ and bulk IGY hard matter state as  $z\rightarrow+\infty$.  The red, blue and brown curves give the profiles of the fermion energy density, the scalar field energy density and the scalar field; all as a function of the distance, $z$, from a nominal center of the phase boundary.  The two horizontal lines give the asymptotic soft and hard matter total energy density.}
\label{fig:PhaseBoundaryProfile}
\end{figure}

\subsubsection{Connecting Hard Matter with Vacuum}
\label{sec:VacuumHardInterface}

The largest coupling, $\geff_f$, for a soft/hard phase boundary is $\geff_\mathrm{vac}$ at which one finds $P=0$, $E_\mathrm{F}=m_f$, $\ed^-_\mathrm{bp}(\geff_\mathrm{vac})=0$ and $\ed^+_\mathrm{bp}(\geff_\mathrm{vac})=\ed^+_0(\geff_\mathrm{vac})$; so this is essentially a phase boundary of hard matter with vacuum.  Hard/vacuum interfaces do not exist for $\geff_f<\geff_\mathrm{vac}$ because $E_\mathrm{F}>m_f$ and hence fermions would flow into the vacuum.  Hard/vacuum interfaces do exist for $\geff_f>\geff_\mathrm{vac}$ because  $E_\mathrm{F}<m_f$ and while fermions would flow from the vacuum there are no fermions in vacuum to flow, i.e. one does not have to have chemical equilibrium with vacuum.  Pressure equilibrium is still required which for vacuum interface is $P=0$.  For IGY these two conditions are only satisfied on the curve $\ed=\ed^+_0(\geff_\mathrm{vac})$ for $\geff_f>\geff_\mathrm{vac}$ which can form a stable interface with vacuum such as on the surface of a star.  These vacuum interface states are shown as the orange dashed curve in fig.~\ref{fig:EoSIGY2} and orange squares in fig.~\ref{fig:InterfaceGraph}.

\begin{figure}
\centering
\includegraphics[width=\textwidth]{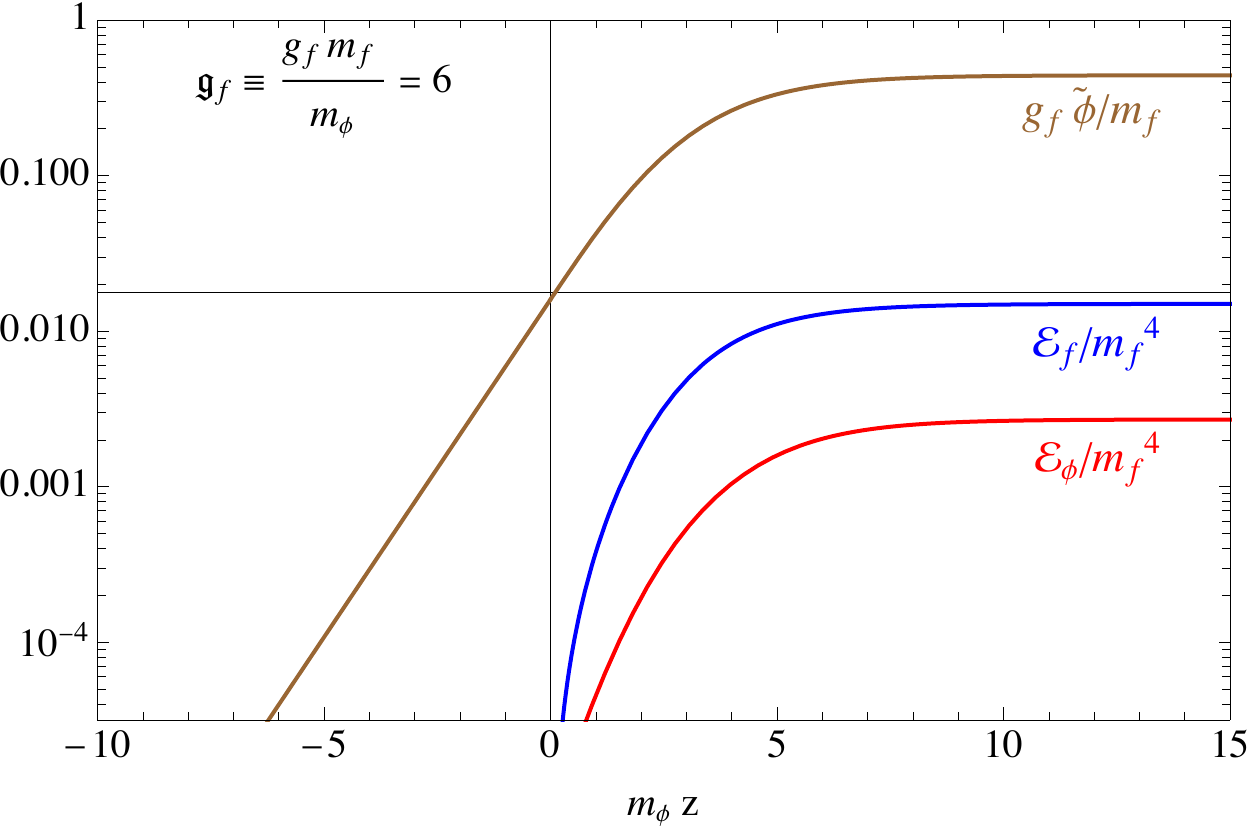}
 \caption{Plotted is the profile of a 1-D phase boundary of vacuum with hard matter for $\geff_f=6$ as described in \S\ref{sec:VacuumHardInterface}.  The blue, red and brown curves give the profiles of the fermion energy density, the scalar field energy density and the scalar polarization; all as a function of the distance, $z$, from the fermion surface.  For  $z\le0$ the fermion energy density, $\ed_f$, is zero while the scalar field falls off $\propto e^{m_\phi\,z}$ and the scalar field energy density falls off $\propto e^{2\,m_\phi\,z}$.   Pure vacuum is reached as $z\rightarrow-\infty$ and bulk hard matter IGY state is reached as $z\rightarrow+\infty$.   The horizontal line gives the total energy density of the asymptotic bulk hard matter.}
\label{fig:VacuumBoundary}
\end{figure}

A microscopic description of the hard/vacuum interface is a 1D hydrostatic configuration described by eq.~\ref{eq:hydrostatic1D} with $P=0$ and $E_\mathrm{F}\le m_f$.  In this case the fermion energy density goes to zero (the argument of $\varphi$ goes to zero) when $\tilde{\phi}=(m_f-E_\mathrm{F})/g_f$.  One can define $z=0$ at this position.  Retaining the $\tilde{\phi}'(z)>0$ convention: for $z>0$ there are fermions while for $z<0$ there are no fermions but $\tilde{\phi}\ne0$.  For $z<0$ eq.~\ref{eq:hydrostatic1D}  reduces to $\tilde{\phi}'(z)=m_\phi\,\tilde{\phi}(z)$ and matching the  $z<0$ and $z>0$ solutions for $\tilde{\phi}$ one finds  $\tilde{\phi}(z)=\frac{m_f-E_\mathrm{F}}{g_f}\,e^{m_\phi\,z}$ for $z\le0$.  In order for the $z>0$ solution to asymptote to a uniform bulk state, $\tilde{\phi}'(+\infty)=0$, one requires that the Fermi energy matches that of the IGY hard matter zero pressure state which is $E_\mathrm{F}=\ed_0^+/n_0^+$.  Thus such an interface has an exponential ``$\phi$ atmosphere'',
\beq
\tilde{\phi}(z<0)=\frac{1}{g_f}\,\left(m_f-\frac{\ed_0^+(\geff_f)}{n_0^+(\geff_f)}\right)\,e^{m_\phi\,z} \qquad n_f(z<0)=0
\label{eq:ExponentialAtmosphere}
\eeq
where $n_0^+$ is the fermion number density when $\ed=\ed_0^+$.  There is no simple expression for the $z>0$ solution but an example numerical solution is shown in fig.~\ref{fig:VacuumBoundary}.  Again we we see that the  transition layer is only a few ${m_\phi}^{-1}$ thick and in the LSA this phase boundary can be treated as discontinuous.

\subsubsection{Two Phase Instability}
\label{sec:TwoPhaseInstability}

These two types of phase boundaries clarify what happens to hydrodynamically unstable cold matter which occurs when $\ed_\sim^-<\ed<\ed_\sim^+$ for $\geff_f>\geff_\mathrm{crit}$.  Consider a thought experiment where one sets up cold Yukawa coupled fermions confined to a finite volume in a uniform unstable state.   Since it is unstable density inhomogeneities will grow releasing free energy.  If one then allows these fermions to cool back to zero temperature, assuming the volume is $\gg{m_\phi}^{-3}$ but not large enough for gravity to play a role, the matter will reach an inhomogeneous two phase equilibrium separated by the phase boundaries described above.  If  $\geff_\mathrm{crit}<\geff_f<\geff_\mathrm{vac}$ then the two phases are  a soft phase with energy density $\ed_\mathrm{pb}^-$ and a hard matter phase with much higher energy density $\ed_\mathrm{pb}^+$.  This is analogous to a liquid with non-zero equilibrium vapor pressure.  If  $\geff_f>\geff_\mathrm{vac}$ the two phases will be vacuum and hard matter with energy density $\ed_0^+$.  This is analogous to a liquid with zero equilibrium vapor pressure.  One could imagine such a two phase instability to occur during the collapse of a cold star.  In this paper we have focused on cold matter but have shown that the Yukawa force decreases as the energy of the particles increases.  Therefore one could also imagine that the onset of this instability could occur as a star cools.

\subsubsection{Metastable Matter}
\label{sec:MetastableMatter}

States with the lowest energy per conserved fermion are energetically preferred. This is determined by the {\it chemical energy}: $\ce_f\equiv\frac{\ed}{n_f}$ which does not include gravitational binding energy.  For an adiabatic EoS the chemical energy extrema are given by $0=\frac{\partial\ce_f}{\partial n_f}=P/n_f^2$, i.e. at $P=0$ points. For IGY these densities are $\ed=0$ and for $\geff_f\ge\geff_0$ also $\ed=\ed_0^\pm$. The chemical energy is minimized, maximized, and minimized at $\ed=0,\,\ed_0^-$ and $\ed_0^+$, respectively.  At these extrema the chemical energy is equal to the chemical potential: $\ce_f=\cp_f$.  The low density minimum is $\ce_f=m_f$ while the high density minimum is $\ce_f=\cp_f=\ed_0^+/n_0^+$.  As indicated above $\ed_0^+<m_f\,n_0^+$ for $\geff_f\ge\geff_\mathrm{vac}$ in which case $\ed_0^+$ is energetically preferred over a low density state.  Since for soft matter $\ce_f$ increases with density $\ed_0^+$ is energetically preferred to {\it all} soft matter for $\geff_f\ge\geff_\mathrm{vac}$. This energetic preference means that energy would be released by transforming soft matter into hard matter.  However the soft matter does not spontaneously transform to this lower energy state since it is hydrodynamically stable. We say that soft matter is {\it metastable} if $\geff_f\ge\geff_\mathrm{vac}$.  Metastability is indicated by yellow in fig.~\ref{fig:EoSIGY2}.  Note that this implies that SM matter would be metastable at low densities if it is coupled to a scalar field with a sufficiently strong Yukawa coupling!

Metastability also occurs when $\geff_\mathrm{crit}<\geff_f<\geff_\mathrm{vac}$ where soft/hard phase boundaries are possible.  At the phase boundary $\ed$ jumps from $\ed_\mathrm{bp}^-$ to $\ed_\mathrm{bp}^+$, jumping over soft, forbidden and hard matter.  All of the denser soft matter in the range $\ed_\mathrm{pb}^-<\ed<\ed_\sim^-$ have larger chemical potentials than those of hard matter with $\ed\ge\ed_\mathrm{pb}^+$ so the fermions would flow from the soft to hard matter if the former came into contact with the latter.  However this denser soft matter is hydrodynamically stable and would not spontaneously transition into hard matter.  In this sense the denser soft matter is also metastable as indicated by yellow in fig.~\ref{fig:EoSIGY2}.

The features of the IGY EoS above the critical point:  soft/forbidden/hard matter, phase boundaries, two phase instabilities and metastable soft phases are generic features of EoSs with an anomalous $P(\ed)$ which rises, falls and then rises again.  

\section{IGY Prototype Yukawa Coupled Neutron Star}
\label{sec:IGYstellarStructure}

Here we compute the structure of a spherical hydrostatic object consisting of matter described by the IGY EoS.  The IGY EoS has relevance to the structure of neutron stars where density and pressure are both dominated by a single fermion species, the neutron, at least over some range of densities.   For this reason we take $f\rightarrow\mathrm{n}$ in this section and quote physical values for quantities based on the neutron mass.  IGY treats baryonic matter as an ideal cold gas of neutron just as was done in the foundational work of Oppenheimer and Volkoff~\cite{PhysRev.55.374} (OV).  The OV model corresponds to an IGY neutron star with $\geff_\mathrm{n}=0$.   IGY and OV have numerous shortcomings in describing neutron stars.  At low densities they do not incorporate the pressure of muons and electrons or the changing abundance of protons and nuclei with pressure.  At higher densities they do not include hadronic interactions which stiffen the EoS; and will eventually lead to the disassociation of neutrons into quarks.  Thus IGY like OV provides only a rough approximation to a neutron star structure over a limited range of densities.  The purpose of this exercise is to illustrate the phenomenology of stellar structure which can be caused by a Yukawa coupling.  The expectation is that qualitative features of IGY {\it prototype} neutron stars will also manifest in stars with general and more realistic EoSs with Yukawa coupling to a light scalar.  

The different phases of IGY matter identified above and the ability to form stable phase boundaries have a profound effect on stellar structure.  This depends on the strength of the Yukawa coupling.  One may define three ranges of $\geff_f$ values according to the allowed phases of matter which hydrostatic configurations may contain:
\begin{itemize}
\item {\bf weak coupling:} $\geff_f\le\geff_\mathrm{crit}=4.987\ldots$
\begin{itemize}
\item[-]  {\it normal stars:} all normal matter
\end{itemize}
\item {\bf moderate coupling:}  $\geff_\mathrm{crit}<\geff_f<\geff_\mathrm{vac}$
\begin{itemize}
\item[-]  {\it soft stars:} all $\ed<\ed_\sim^-$ soft matter 
\item[-]  {\it hybrid stars:} outer $0<\ed<\ed_\mathrm{pb}^-$ soft shell with inner $\ed>\ed_\mathrm{pb}^+$ hard core
\end{itemize}
\item {\bf strong coupling:} $\geff_f\ge\geff_\mathrm{vac}=5.681\ldots$
\begin{itemize}
\item[-]  {\it soft stars:} all $\ed<\ed_\sim^-$ soft matter
\item[-]  {\it hard stars:} all $\ed>\ed_0^+$ hard matter 
\end{itemize}
\end{itemize}
The densities referred to here are in the macroscopic light scalar approximation, LSA, not the microscopic description illustrated in fig.s~\ref{fig:PhaseBoundaryProfile}~\&~\ref{fig:VacuumBoundary}.  IGY hybrid stars are analogous to hybrid quark stars \cite{Rosenhauer1991} which have have been extensively studied.

Since the IGY has isotropic pressure, $P(\ed)$, the spherical stellar structure, $\ed(r)$, is given by solutions to eq.~\ref{eq:TOVequation} with boundary conditions $\ed'(0)=0$ and $P(\ed(R_\star))=0$.  Zero surface pressure requires that $\ed(R_\star)=0$ for all solutions {\it except} hard stars which have $\ed(R_\star)=\ed_0^+$.  Solutions can be parameterized by the effective Yukawa coupling, $\mathfrak{g}_\mathrm{n}$, and the energy density at the center, $\ed(0)$.

We have determined the stellar structure, $\ed(r)$, $P(r)$, $M_\star$, $R_\star$, etc. for thousands of IGY stellar models parameterized by $\geff_\mathrm{n}$ and $\ed(0)$.  Below we will only present a sample of these results which we believe is large enough for the reader to get a sense of the phenomenology to be able to interpolate in this parameter space.  We also briefly explain why the solutions behave as they do.  We begin with some gross features of the solutions and then go into more details.

Not all hydrostatic solutions will be stable which further limits viable hydrostatic models of stars.  The hydrodynamical instability identified in \ref{sec:ForbiddenSoftHard} is for a uniform medium neglecting gravity.  Stars are not uniform and may have other hydrodynamic or gravitational instabilities.  Below we treat such global instabilities in only a perfunctory way by determining its {\it nominal stability} according to the sign of $\frac{d M_\star}{d\ed(0)}$ \cite{Bardeen1966}.  Negative values are indicative of instability to gravitational core collapse.  A definitive determination of stability by normal mode analysis is considered too cumbersome to apply to the large parameter space of stellar models under consideration.

\subsection{Mass and Radius}
\label{sec:MassRadius}

The most easily determined (measured) quantity for observed neutron stars is their rotation rates.  However this does not bear much relation to the star's EoS so it is not most relevant for this paper.  A more relevant quantity is the gravitational mass of neutron stars, $M_\star$, which has been determined with some accuracy for a large number of stars.  $M_\star$ combined with direct measurement of neutron star radii, $R_\star$ (currently few in number) will provide strong constraints on the neutron stars EoS. Gravitational wave observations of merging neutron stars can also measure combined tidal deformablilty which depends on the density profile within the star with a strong dependence on $R_\star$.  A complete $M_\star(R_\star)$ relation (accounting for rotation) would determine the EoS $\ed(P)$ if, as expected, the nuclear matter EoS is universal.  A theoretical $M_\star(R_\star)$ curve (usually) gives a maximal mass,  $M_\star^\mathrm{max}$, and minimum radius, $R_\star^\mathrm{min}$, which are generally achieved simultaneously in the {\it maximal star}.  Prediction of these quantities are of particular interest as it allows one to exclude candidate EoSs using single star measurements of either $M_\star$ or $R_\star$.

\begin{figure}
\centering
\includegraphics[width=\textwidth]{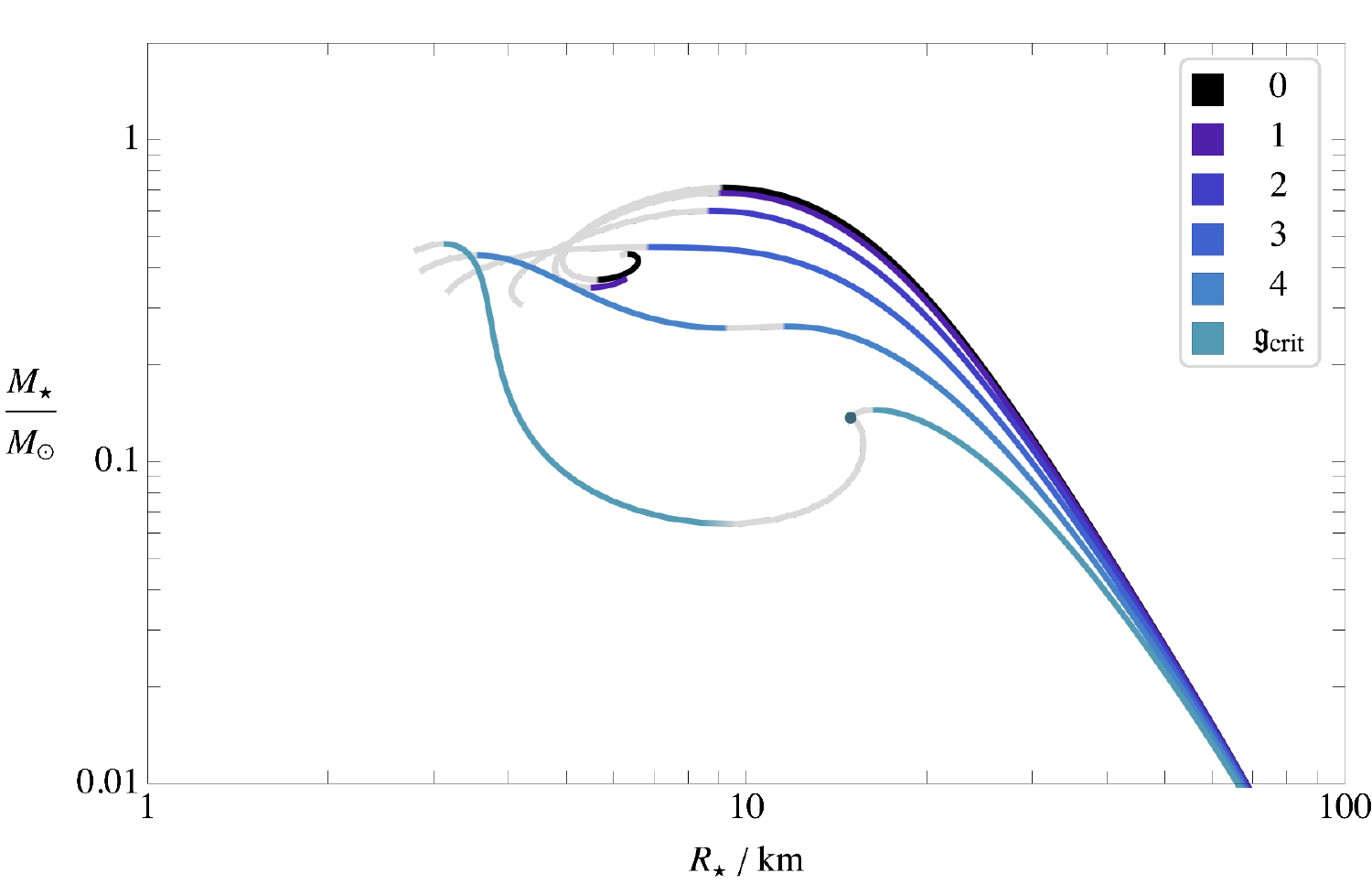}
 \caption{$M_\star$ vs. $R_\star$ for spherical hydrostatic configurations of matter with the IGY equation of state for the values of the effective coupling, $\geff_\mathrm{n}\equiv\frac{g_\mathrm{n}\,m_\mathrm{n}}{m_\phi}$, as indicated in the legend.  The couplings in this graph are limited to the range of weak couplings $0\le\geff_\mathrm{n}\le\geff_\mathrm{crit}$.  $M_\star$ is the total gravitational mass and $R_\star$ the circumferential radius. Solutions which are nominally gravitationally unstable are plotted in light gray.  The dot gives the critical star solution where  where $\geff_\mathrm{n}=\geff_\mathrm{crit}$ and the central density is $\ed(0)=\ed_\mathrm{crit}$ where $P'(\ed)=P''(\ed)=0$.  This unstable solution is the separatrix between normal stars which are illustrated here and the soft and hybrid stars which occur for larger coupling.
}
\label{fig:MofRweakIGY}
\end{figure}

\begin{figure}
\centering
\includegraphics[width=\textwidth]{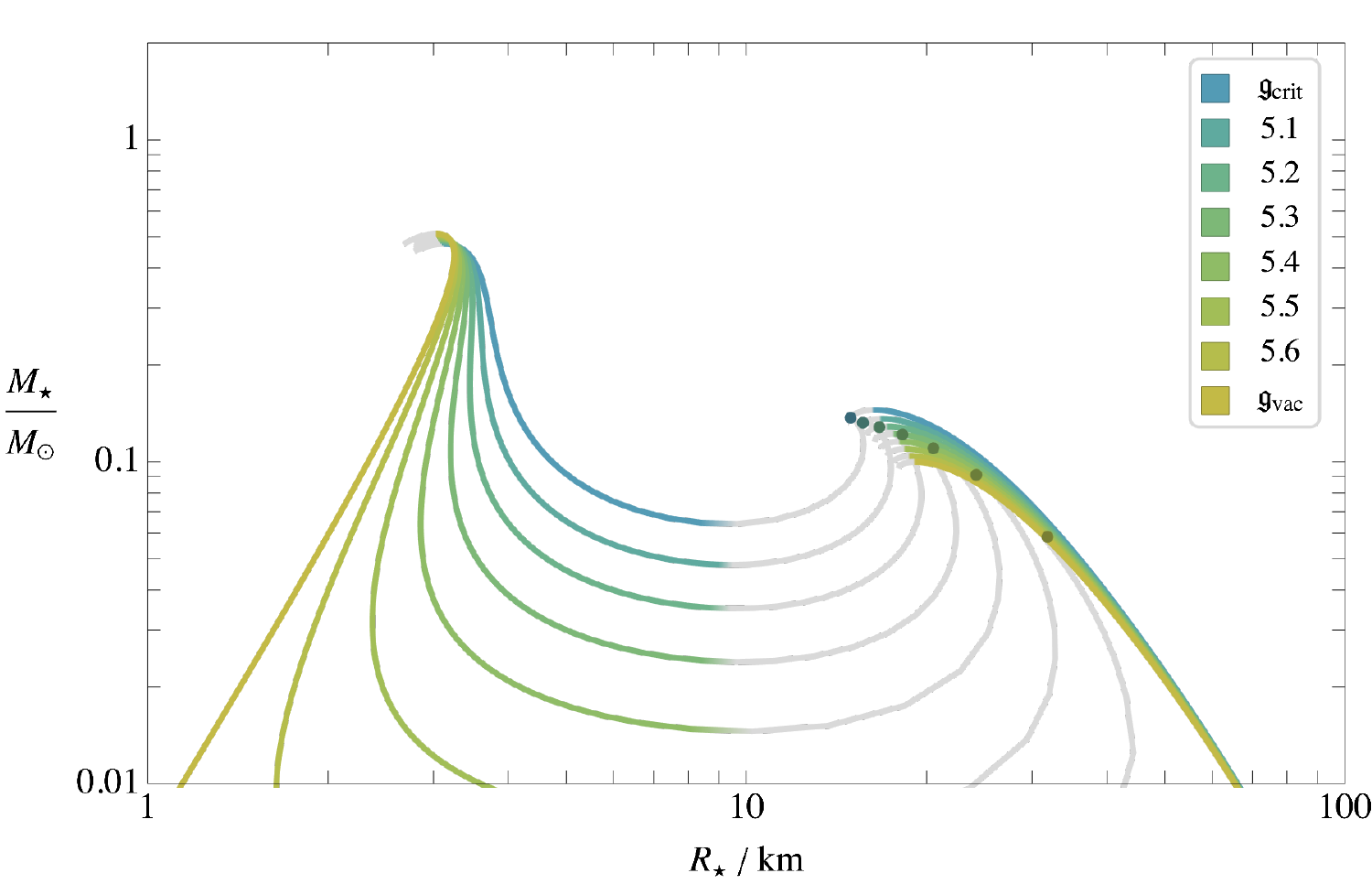}
 \caption{$M_\star$ vs. $R_\star$ with the IGY equation of state in the range of moderate couplings $\geff_\mathrm{crit}\le\geff_\mathrm{n}\le\geff_\mathrm{vac}$. There are two branches of the $M_\star(R_\star)$ solutions which coincide at bifurcation solutions denoted by the dots .  The larger radius $M_\star(R_\star)$ curve consists only of stars made of soft matter which we call {\it soft stars}.  The other curve emanating from the dot (which resembles the outline of a swan) are {\it hybrid stars} which consist of hard matter in the center and soft matter on the outside joined together by the soft/hard interface described in \S\ref{sec:SoftHardInterface}.  The gray parts of the curve are solutions which are nominally gravitationally unstable.
}
\label{fig:MofRmoderateIGY}
\end{figure}

\begin{figure}
\centering
\includegraphics[width=\textwidth]{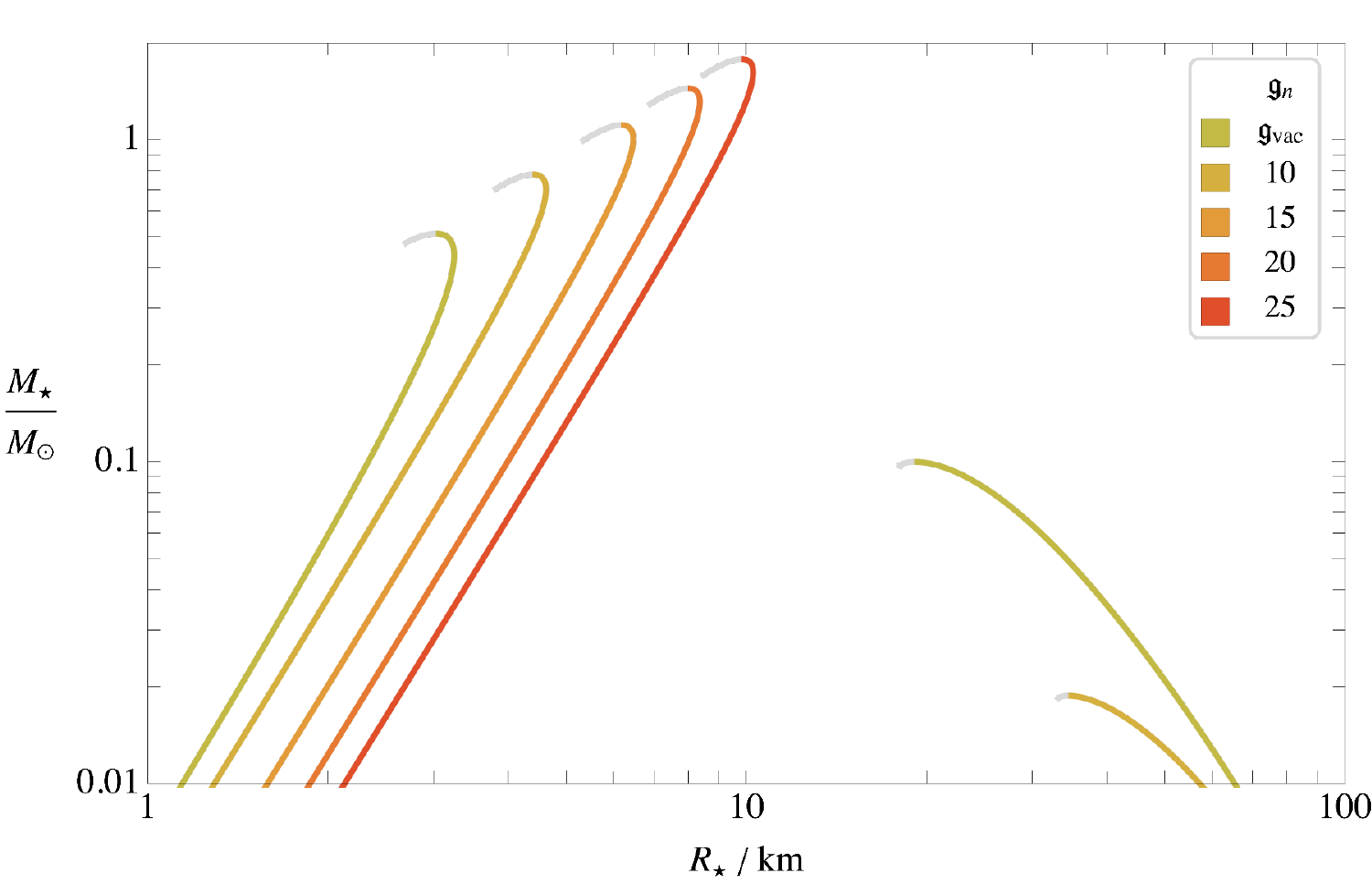}
 \caption{$M_\star$ vs. $R_\star$ with the IGY equation of state in the range of strong couplings $\geff_\mathrm{n}\ge\geff_\mathrm{vac}$. There are two separated branches of the $M_\star(R_\star)$ solutions.  The curves on the right  {\it soft star} solutions consisting only of soft matter.  The curves on the left (which resemble canes) are {\it hard stars} which consist only of hard matter and are joined on the surface to vacuum by the vacuum/hard interface described in \S\ref{sec:VacuumHardInterface}.  The gray parts of the curve are solutions which are nominally gravitationally unstable.
}
\label{fig:MofRstrongIGY}
\end{figure}

With this motivation we start the exposition of solutions to the TOV equation for IGY stars by plotting $M_\star(R_\star)$ for various values of  $\geff_\mathrm{n}$ while taking the fermion to be a neutron.  For clarity we plot $M_\star(R_\star)$ for weak, moderate and strong coupling separately; in fig.s~\ref{fig:MofRweakIGY}, \ref{fig:MofRmoderateIGY} \& \ref{fig:MofRstrongIGY}, respectively.  The black $\geff_\mathrm{n}=0$ curve in fig.~\ref{fig:MofRweakIGY} is the OV model which is the ``standard model'' curve with which to compare the other curves.  Irrespective of  $\geff_\mathrm{n}$ all IGY models at small masses only sample the $\Gamma\equiv\frac{d\mathrm{ln}P}{d\mathrm{ln}\ed}\approx\frac{5}{3}$ part of the EoS in fig.~\ref{fig:EoSIGY1}.  Thus IGY $M_\star(R_\star)$ asymptotes to a $\Gamma=\frac{5}{3}$ polytrope (polytropic index $\frac{3}{2}$) for small masses / low central densities where $M_\star\propto R_\star^{-3}$.  These are the stellar models corresponding to the points on the curves extending to the lower right in fig.s~\ref{fig:MofRweakIGY},~\ref{fig:MofRmoderateIGY}~\&~\ref{fig:MofRstrongIGY}.  It is $M_\star(R_\star)$ where the EoS deviates from this polytropic limit which depends on $\geff_\mathrm{n}$. 
    
Most EoSs, including IGY, asymptote to ultra-relativistic matter with $\Gamma\rightarrow1$, i.e. become softer, at large enough density.  Such a small $\Gamma$ at the center of a star indicates instability to core collapse.  Thus there is a maximum central density, $\ed_\mathrm{max}$, for stable stellar solutions and the star central density $\ed_\mathrm{max}$ yields the maximal star with maximum mass, $M_\star^\mathrm{max}$.  As can be seen from fig.~\ref{fig:EoSIGY1} for $\geff_\mathrm{n}\gtrsim4$ the EoS can also become soft at intermediate densities.  This soft EoS at intermediate densities leads to nominally unstable solutions (plotted in gray) spanning a range of $R_\star$ for $4\lesssim\geff_\mathrm{n}<\geff_\mathrm{vac}$ in fig.s~\ref{fig:MofRweakIGY}~\&~\ref{fig:MofRmoderateIGY}.

For weak couplings, $\geff_\mathrm{n}<\geff_\mathrm{crit}$, since $\Gamma>0$ one can obtain the pressure gradient, $P'(r)$, required by eq.~\ref{eq:TOVequation} to support the star with a sufficiently large density gradient, $\ed'(r)$.  For moderate couplings this is not always possible since at intermediate densities $\Gamma<0$ for $\geff_\mathrm{n}>\geff_\mathrm{crit}$.  As one transitions from the weak coupling stellar solutions in fig.~\ref{fig:MofRweakIGY} to the moderate coupling solutions in fig.~\ref{fig:MofRmoderateIGY} these large density gradients are replaced by the discontinuous soft/hard phase interface described in \S\ref{sec:SoftHardInterface}.  For moderate couplings there is not only an instability gap between large and small radius solutions but there are two separate branches of solutions to the the TOV equation.  In fig.~\ref{fig:MofRmoderateIGY} these two branches bifurcate at a stellar solution indicated by a dot which are stars with central density $\ed(0)=\ed_\mathrm{pb}^-(\geff_\mathrm{n})$. The larger $R_\star$ branch are {\it soft stars} consisting only of soft matter. The soft matter branch extends beyond the bifurcation point to larger mass and smaller radius.  In this extension the center consists of chemically metastable soft matter described in \S\ref{sec:MetastableMatter}.  The other branch extending from the bifurcation point, tracing the outline of a swan, are {\it hybrid stars} consisting of hard matter in the center and soft matter on the outside separated by a soft/hard interface.  This interface is at the center of the star at the tip of the swan's tail and moves outward along this curve but never reaches the surface. The swan's tail region is nominally gravitationally unstable but becomes stable in the swan's back at lower mass and radius proceeding up the neck of the swan to the top of it's head where the solution again becomes gravitationally unstable at the maximal star.

As $\geff_\mathrm{n}\rightarrow\geff_\mathrm{vac}$ from below the bifurcation point moves to $M_\star\rightarrow0$ and  $R_\star\rightarrow\infty$.  The swan's back and tail also move to $M_\star\rightarrow0$ and disappear leaving only the head and the neck.  For strong coupling, $\geff_\mathrm{n}>\geff_\mathrm{vac}$, there is no stable hard/soft interface, so hybrid stars cannot exist, and there is no bifurcation point.  Instead the two solution branches separate into two disconnected solutions as shown in fig.~\ref{fig:MofRstrongIGY}.  A gap exists between these two solutions but not because of gravitational instability; rather there are no hydrostatic solutions, stable or unstable, in this gap.  The large $R_\star$ solutions are soft stars consisting solely of soft matter.  Soft star solutions as seen for moderate couplings continue into the hard coupling regime.  The maximum soft star mass shrinks with $\geff_\mathrm{n}$, increasing the gap between the two branches, since the energy density becomes softer at lower density for larger $\geff_\mathrm{n}$ as can be seen from fig.~\ref{fig:EoSIGY1}.  The small $R_\star$ solutions are {\it hard stars} consisting solely of hard matter with a vacuum interface on the surface as described in \S\ref{sec:VacuumHardInterface}.  The hard star $M_\star(R_\star)$ curves are cane shaped: the swan's neck morphed into the length of the cane and the swan's head into the cane's handle. The top of the handle is the maximal star while the length of the cane below the handle describe constant density spheres with $M_\star\propto{R_\star}^3$ extending to very low mass and radius.  We call solutions in the constant density regime {\it hard drops} and discuss them further in \S\ref{sec:HardDrops}.  Hard stars fill the $R_\star$-$M_\star$ plane approximately up to a maximum density giving $M_\star\lesssim0.0064\,\left(\frac{R_\star}{\mathrm{km}}\right)^3\,M_\odot$ and a maximum surface gravitational potential giving $M_\star\lesssim0.2\,\frac{R_\star}{\mathrm{km}}\,M_\odot$ (see \S\ref{sec:HardDrops} \& \S\ref{sec:MaximalStars}).  For any normal, soft or hybrid star there will be a hard star with larger $\geff_\mathrm{n}$ but the same $M_\star$ and $R_\star$.

\subsection{Hard Drops}
\label{sec:HardDrops}

{\it Hard drops} form the low mass end of the hard star solutions for strong coupling, $\geff_\mathrm{n}\ge\geff_\mathrm{vac}$.  Hard drops are not primarily bound by gravity but rather by the attractive $\phi$ mediated force between neutrons.  The structure of hard drops is simple: on the surface there is a vacuum/hard interface as described in \S\ref{sec:VacuumHardInterface} while in the interior there is uniform hard matter.  Since gravity is unimportant the hard matter is in hydrostatic equilibrium with zero pressure vacuum and has nearly\footnote{Similarly to a water drop surface tension and finite size leads to an interior pressure slightly in excess of the ambient pressure.} zero pressure itself, so the energy density takes the value $\ed_0^+$.  Thus each of these solutions have roughly constant energy density throughout the star: $\ed(R)\approx\ed_0^+$, so $M_\star\approx\frac{4\pi}{3}\,\ed_0^+\,{R_\star}^3$.   Hard drop sizes extend down to the range of the $\phi$-force, $R_\star\gtrsim{m_\phi}^{-1}$, which is the range of validity of the LSA (light scalar approximation).  Depending on $m_\phi$ the minimum hard drop size can be very much smaller than the $\sim10\,$km size of neutrons stars. To $<15\%$ accuracy we find for IGY hard drops the density and chemical energy are $\ed_0^+\approx2\,\frac{\geff_\mathrm{n}-4}{{\geff_\mathrm{n}}^3}\,{m_\mathrm{n}}^4$ and 
$\ce_\mathrm{n}\equiv\frac{\ed_0^+}{n_0^+}\approx\sqrt{\frac{\geff_\mathrm{vac}}{\geff_\mathrm{n}}}\,m_\mathrm{n}$.  Thus the mass-radius relation is $M_\star\approx\frac{8\pi\,(\geff_\mathrm{n}-4)}{3\,{\geff_\mathrm{n}}^3}\,{m_\mathrm{n}}^4\,{R_\star}^3$ up to the maximum density
$\frac{3\,M_\star}{{4\pi\,R_\star}^3}<\frac{\ed_\mathrm{vac}}{n_\mathrm{vac}}\approx\frac{{m_\mathrm{n}}^4}{60}\approx 0.0015\,\,M_\odot/\mathrm{km}^3$ when $\geff_\mathrm{n}=\geff_\mathrm{vac}$.  The binding energy per neutron is 
$m_\mathrm{n}-\ce_\mathrm{n}\approx m_\mathrm{n}\,\left(1-\sqrt{\frac{\geff_\mathrm{vac}}{\geff_\mathrm{n}}}\right)$ which can be a large fraction $m_\mathrm{n}$.

The hard drop name derives from the similarity to drops of a normal liquid which consist of atoms or molecules bound together by local forces in hydrostatic equilibrium with the surrounding medium.  Hard drops like normal drops will have a surface tension which adjusts the surface to minimize its curvature but as this is only a surface effect any configuration with the same volume, whether in different shapes or in multiple drops, have nearly identical energy.  The gravitational force is long range and although it plays a subdominant role in binding a hard drop it may nevertheless be the dominant force in keeping the drop spherical and in one piece; or making it aspherical in the presence of tidal forces.

Small hard drops are not sufficiently massive to be candidates for observed neutron stars but one might imagine that the debris field produced during neutron star formation or neutron star mergers might contain such drops.  A hard drop thus produced would form a seed for conversion to hard matter of any low density matter it comes into contact, since for strong couplings the soft matter is always metastable to conversion to hard matter.  Since the binding energy released per baryon is very large one would expect such a conversion to produce observable phenomena.  Accretion of the interstellar medium onto hard drops would cause them to glow.  Interaction with stars or planets could release an enormous amount of energy, however since the radiation emitted would be Eddington limited, $L_\gamma<3\times10^4\,\frac{M}{M_\odot}\,L_\odot$, they would not be exceedingly luminous.  Super-Eddington luminosity would likely blow away much of the stellar matter before it could convert.

The fact that the dark matter density is only a factor of $\sim5\times$ larger than the baryon density is suggestive that these two cosmological remnants are closely related.  One should consider hard drops (possibly of quarks rather than nucleons) as a candidate for dark matter since then the dark and baryonic matter would have a common origin in a cosmological baryon asymmetry.  Hard drops would carry most of this asymmetry but in a sequestered form.  How hard drops could form in the early universe is a matter of speculation.  Hard drops and stars share a number of similarities to strange quark nuggets \cite{Witten1984}  and stars \cite{Alcock1986}, preon stars \cite{Hansson2005} and neutrino nuggets \cite{AfshordiZaldarriagaKohri2005}.

\subsection{Radius Gap}
\label{sec:RadiusGap}

\begin{figure}
\centering
\includegraphics[width=\textwidth]{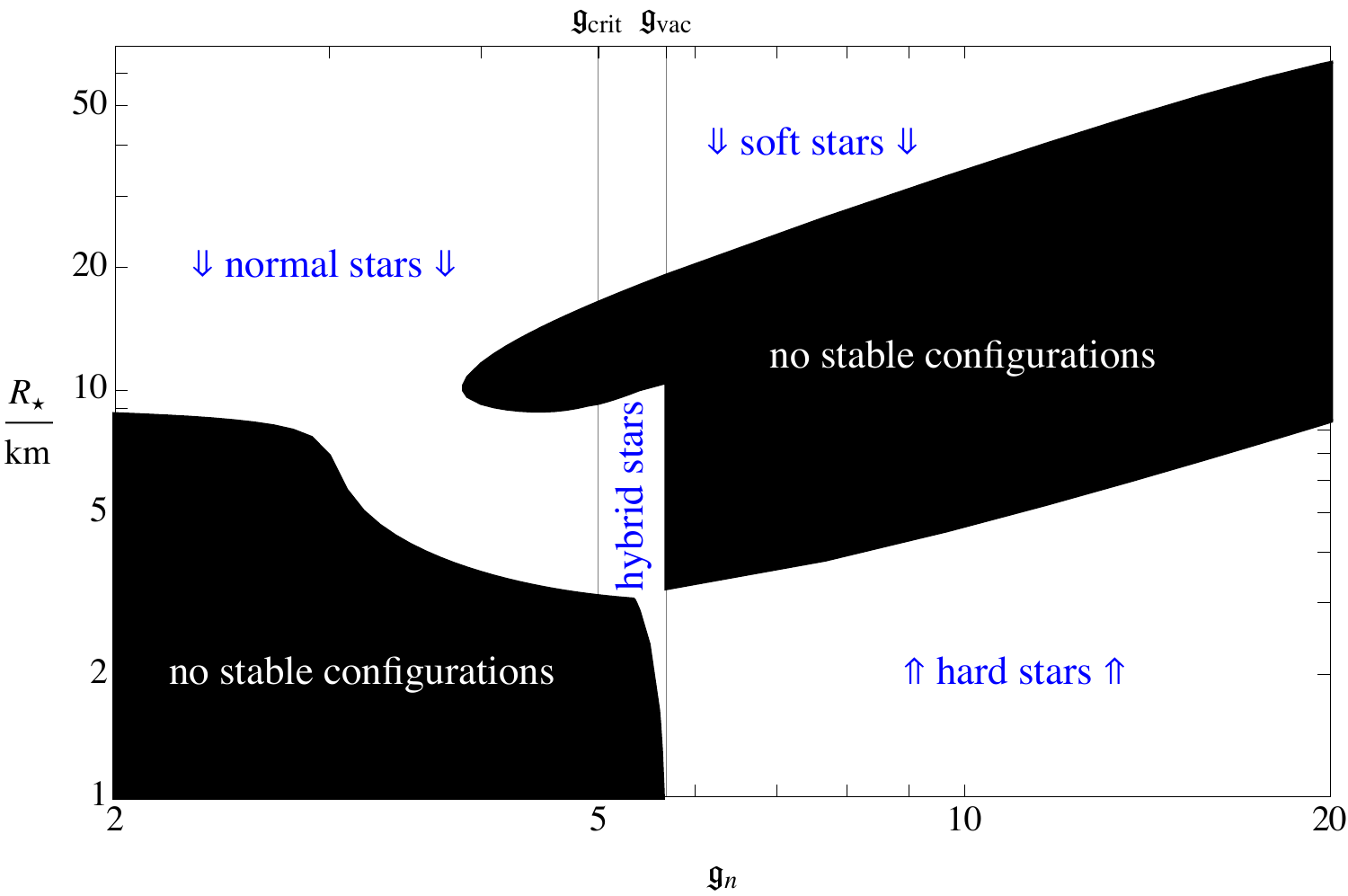}
\caption{
Plotted as a function of the coupling $\geff_\mathrm{n}\equiv\frac{g_\mathrm{n}\,m_\mathrm{n}}{m_\phi}$ is the range of stellar radii, $R_\star$, of nominally stable solutions to the TOV equation given the IGY EOS. The IGY EoS is an idealization and the excluded regions plotted here (in black) will vary for a more realistic EoS though we expect the qualitative geometry of the excluded regions to be insensitive to improvements in the EoS.  The condition for (nominal) stability used here is $\frac{d M_\star}{d\ed(0)}>0$.  The labels give the different classes of stars as described in the text.  The arrows point in the ``general'' direction of increasing $M_\star$.  This directionality cannot be made precise because multiple solutions with the same $R_\star$ may have different $M_\star$ and different signs of $\frac{d M_\star}{d R_\star}$.
}
\label{fig:IGYRadiusExclusion}
\end{figure}

For $\geff_\mathrm{n}\gtrsim4$ IGY stars have a gap in the allowed radii.   In this radius range solutions to the TOV equations are either unstable or do not exist.  A gap can exist for weak coupling normal stars (fig.~\ref{fig:MofRweakIGY}) and always exists for moderate coupling hybrid stars (fig.~\ref{fig:MofRmoderateIGY}).  The gap also always exists for strong coupling between soft and hard stars (fig.~\ref{fig:MofRstrongIGY}).  As can be seen from these figures radius gaps are accompanied by stable solutions with double valued $R_\star(M_\star)$.  Fig.~\ref{fig:IGYRadiusExclusion} shows the gap as a function of $\geff_\mathrm{n}$.  The radius gap appears at $\geff_\mathrm{n}\gtrsim4$ and spans nearly an order of magnitude in radius for $\geff_\mathrm{n}>\geff_\mathrm{vac}$.

\subsection{Maximal Stars}
\label{sec:MaximalStars}

The maximum measured neutron star mass has provided an important constraint on nuclear EoSs.  The theoretical maximal star, the spherical hydrostatic configuration with the maximum mass, is a local maximum of $M_\star(R_\star)$, defining a boundary between the nominally stable and unstable solutions.   For IGY stars one can identify these points on the curves in fig.s~\ref{fig:MofRweakIGY},~\ref{fig:MofRmoderateIGY},~\&~\ref{fig:MofRstrongIGY}: the maximal star is at the top of the swan's head for moderate coupling and the top of the cane's handle for strong coupling.  Where there is a radius gap the maximal star lies below the gap.  Denote the maximal star's mass and radius by $M_\star^\mathrm{mx}$ and $R_\star^\mathrm{mx}$.  Adjacent solutions with $R_\star<R_\star^\mathrm{mx}$ are nominally unstable and solutions with $R_\star>R_\star^\mathrm{mx}$ are nominally stable.

\begin{figure}
\centering
\includegraphics[width=\textwidth]{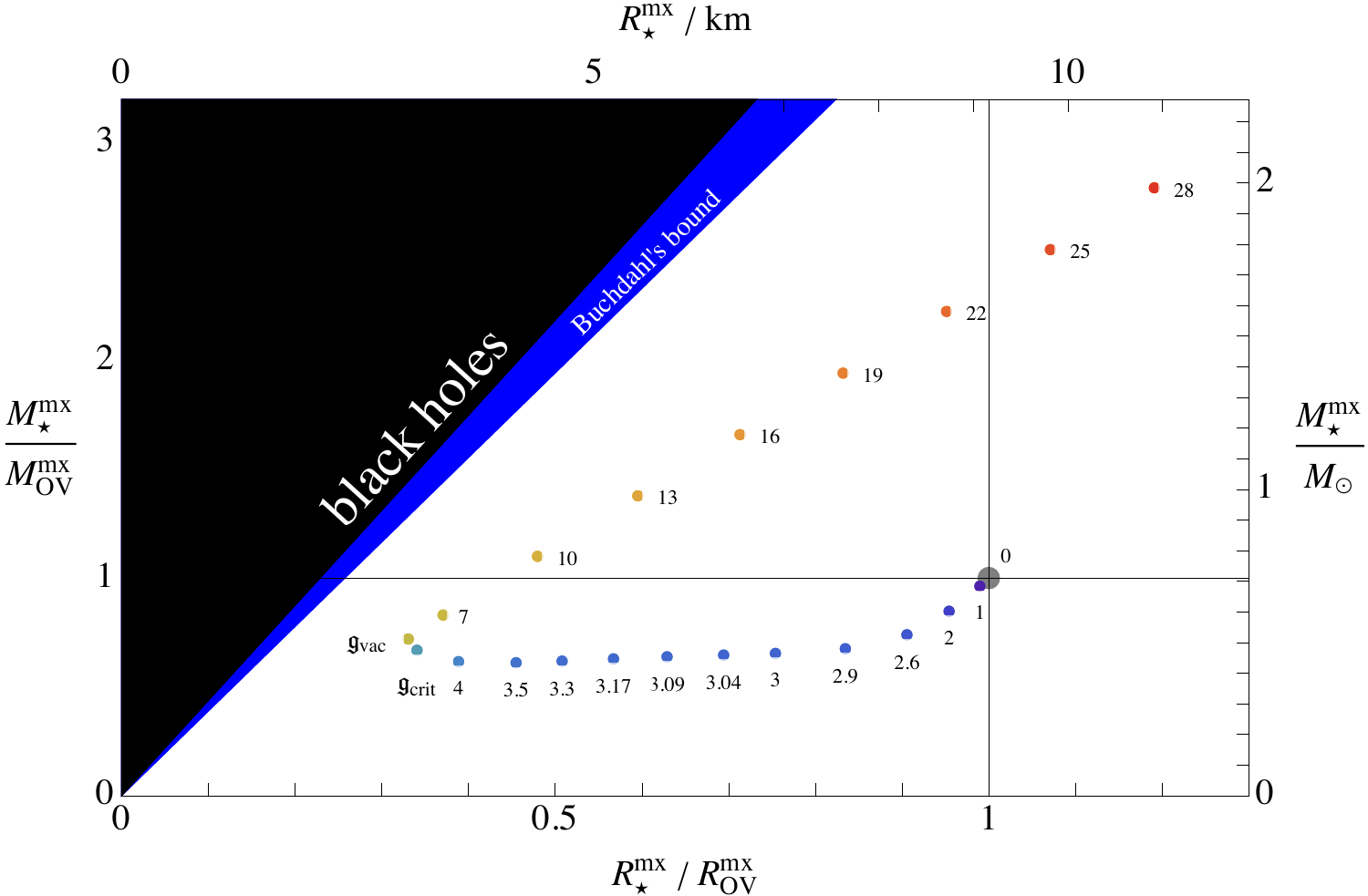}
\caption{
Plotted  for various $\geff_\mathrm{n}$ is the radius and mass of the maximal IGY star, the spherical hydrostatic solution configuration with maximal mass.  Each points is labelled by $\geff_\mathrm{n}=\frac{g_\mathrm{n}\,m_\mathrm{n}}{m_\phi}$.  Note $\geff_\mathrm{crit}\approx4.987$ and  $\geff_\mathrm{vac}\approx5.681$.  To emphasis the change in stellar properties caused by non-zero Yukawa couplings the left and bottom axes give mass and radius in relative to that of the $\geff_\mathrm{n}=0$ OV star represented by the large gray dot.  The right and top axes give mass in radius in physical units using $m_f=m_\mathrm{n}$ appropriate for a neutron star.  None of these models account for nuclear interactions which would increase both the mass and the radius of a more realistic neutron star.  The blue and black shaded regions are disallowed: $\frac{G\,M_\star}{R_\star}\ge\frac{4}{9}$ and $\frac{G\,M_\star}{R_\star}\ge\frac{1}{2}$.  The former is an upper limit derived for an infinitely stiff EoS \cite{Buchdahl1959} and the latter are black holes.
}
\label{fig:MRmaximalIGY}
\end{figure}

Fig.~\ref{fig:MRmaximalIGY} plots $M_\star^\mathrm{mx}$ and $R_\star^\mathrm{mx}$ for selected values of $\geff_\mathrm{n}$ from $0$ to $27$.  Signification deviations from the maximal OV star occurs only for $\geff_\mathrm{n}>1$; for $\geff_\mathrm{n}\lesssim5$ the Yukawa coupling results in a decrease in $M_\star^\mathrm{mx}$ of up to $\sim40\%$ and a decrease in  $R_\star^\mathrm{mx}$ by up to $\sim70\%$.  This trend is rapidly reversed at $\geff_\mathrm{n}\sim5$ where the maximal star's mass and radius begins to rise.  For $\geff_\mathrm{n}>6$  the maximal star has $\frac{M_\star^\mathrm{mx}}{M_\odot}\approx\frac{R_\star^\mathrm{mx}}{5\,\mathrm{km}}\approx\frac{\geff_\mathrm{n}}{12}$ which can be compared with the OV maximal star with $M_\star^\mathrm{mx}\approx0.7\,M_\odot$ and  $R_\star^\mathrm{mx}\approx9\,$km.  Maximal hard stars exceeds the OV mass for $\geff_\mathrm{n}\gtrsim10$ while the maximal hard star radius exceeds the OV value only for $\geff_\mathrm{n}\gtrsim23$.  Depending on $\geff_\mathrm{n}$ maximal IGY stars may be more or less massive than the  maximal OV star and larger or smaller than the maximal OV star.  No maximal IGY star is both less massive and larger in size than the maximal OV star.  $\frac{G\,M_\star^\mathrm{mx}}{R_\star^\mathrm{mx}}\approx0.26$ for maximal hard stars in comparison to 0.114 for the maximal OV star and $\frac{4}{9}$ which is Buchdahl's bound on this quantity \cite{Buchdahl1959}.

\subsection{Density Profiles}
\label{sec:DensityProfiles}

\begin{figure}
\centering
\includegraphics[width=\textwidth]{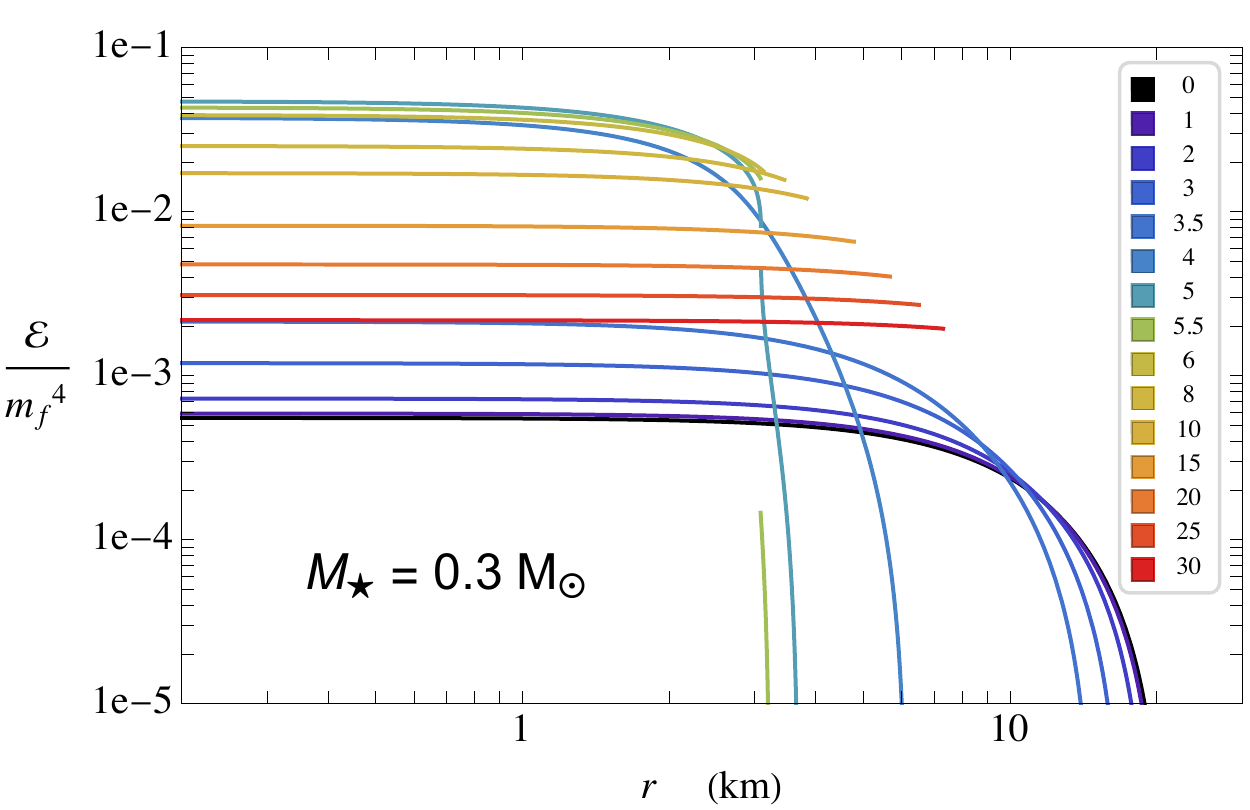}
\caption{
The radial density profile, $\ed(r)$, for IGY stars of gravitational mass $M_\star=0.3\,M_\odot$ with values of the $\geff_\mathrm{n}=\frac{g_\mathrm{n}\,m_\mathrm{n}}{m_\phi}$ as indicated in the legend.  For $\geff_\mathrm{n}=0,\,\ldots,\,4$ these are normal stars.  For  $\geff_\mathrm{n}=5,\,5.5$ these are hybrid stars with the density discontinuity inside the star illustrated by a gap in the $\ed(r)$ curve.  For  $\geff_\mathrm{n}=6,\,\ldots,\,30$ these are hard stars with a discontinuous density drop to zero (vacuum or $\ed=0$) at the end of the curves.   Density discontinuity is a macroscopic description; see \S\ref{sec:Interfaces} for microscope description.   For $\geff_\mathrm{n}>30$ the uniform density hard drop description of \S\ref{sec:HardDrops} is accurate.
}
\label{fig:DensityProfiles03SM}
\end{figure}

To illustrate the different types of density profiles IGY predicts in  fig.~\ref{fig:DensityProfiles03SM} is plotted the radial density profile for stars all of the same mass $M_\star=0.3\,M_\odot$ but for different $\geff_\mathrm{n}$.  For this $M_\star$ there is only a single stable configuration for all $\geff_\mathrm{n}$ which would not be true of smaller $M_\star$.  Shown are four normal stars with weak coupling, two hybrid stars with moderate  coupling and seven hard stars with strong coupling. The central density, $\ed(0)$, increases with $\geff_\mathrm{n}$ by $\sim100\times$ up to moderate couplings of $\geff_\mathrm{n}\sim5$, and then decreases.  The stellar radius, $R_\star$, decreases as $\ed(0)$ increases as one would expect.  In the IGY model $\ed(0)\rightarrow0$ and $R_\star\rightarrow\infty$ as $\geff_\mathrm{n}\rightarrow\infty$.  As one increases $\geff_\mathrm{n}$ the profile approaches a step function: $\ed(r)=\ed(0)\,\Theta(R_\star-r)$ as one obtains for hard drops. For the hybrid stars illustrated the density decrease at the soft/hard interface is $\sim2\times$ and $\sim100\times$ for $\geff_\mathrm{n}=5$ and $\geff_\mathrm{n}=5.5$, respectively.  The hard matter core extends out to 83.7\% and 95.6\% of the stellar radius and contains 98.4\% and 99.97\% of the stellar mass for these two couplings,  The soft matter envelope not only has small densities but steeply falling densities. The density inside hard stars does not vary more than a factor of 2 from it's surface value $\ed(R_\star)=\ed_0^+$ and is very nearly constant for $\geff_\mathrm{n}\gg\geff_\mathrm{vac}$ which is the hard drop limit.

\subsection{Binding Energy}
\label{sec:BindingEnergy}

\begin{figure}
  \centering
    \centering
\vskip-20pt
    \subcaptionbox{Rest mass ($\equiv m_\mathrm{n}\,N_\mathrm{b}$).}
      {\includegraphics[width=0.8\textwidth]{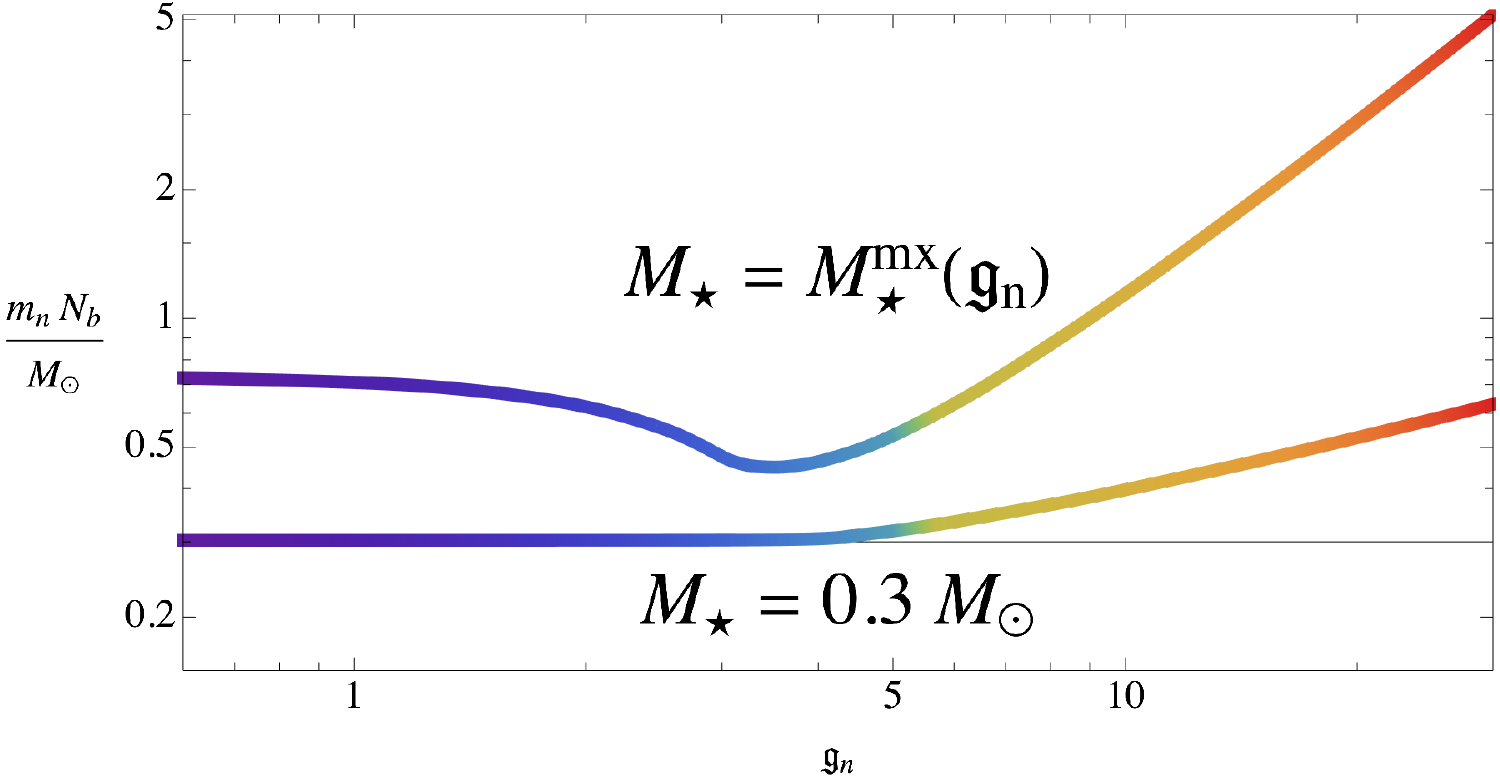}}
\vskip10pt
    \subcaptionbox{Average binding energy per baryon ($b_\star\equiv\frac{m_\mathrm{n}\,N_\mathrm{b}-M_\star}{N_\mathrm{b}}$).}
      {\includegraphics[width=0.8\textwidth]{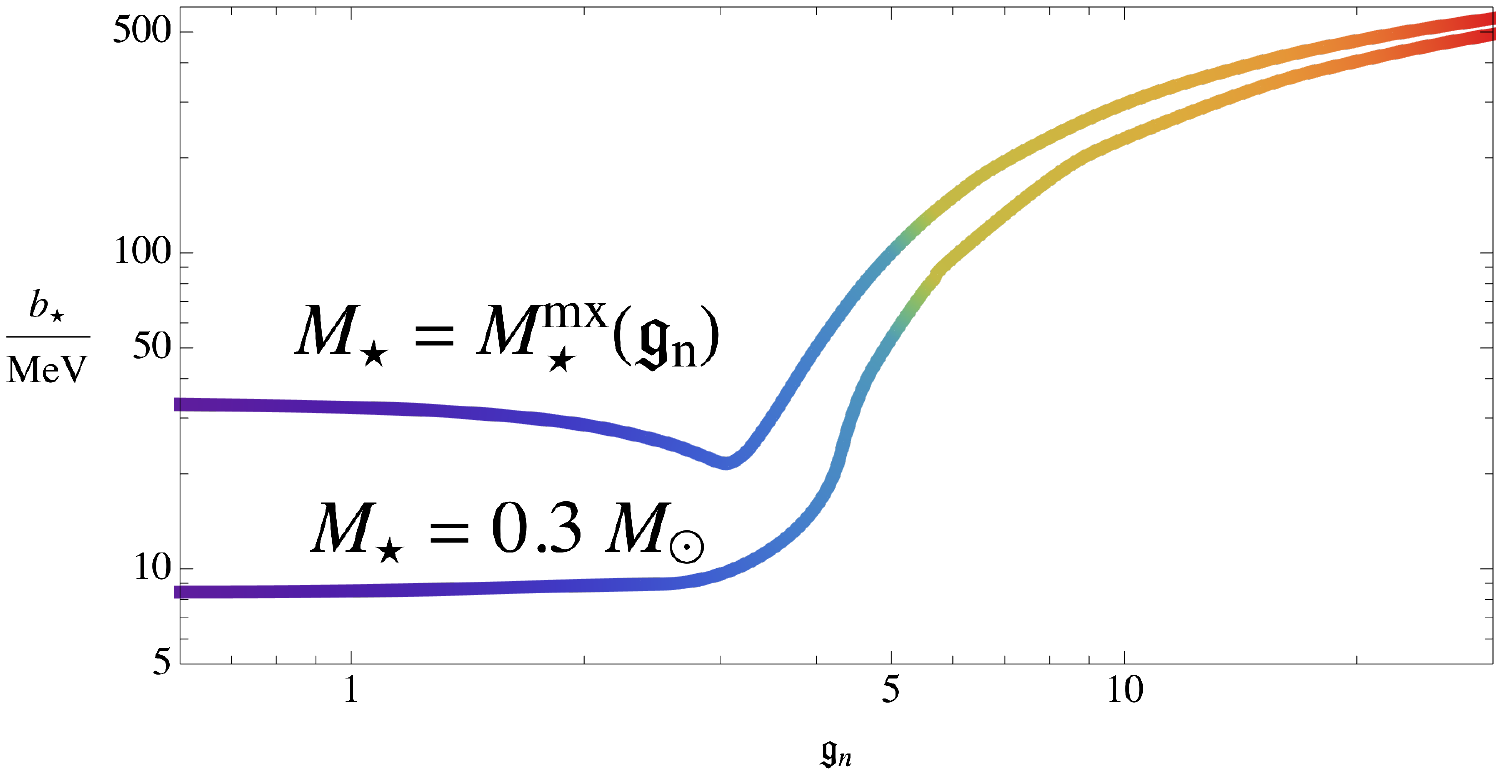}}
      \label{fig:BindingEnergy}
\vskip10pt
    \subcaptionbox{Energy released during formation ($E_\mathrm{rad}\approx b_\star\,N_\star$).}
      {\includegraphics[width=0.8\textwidth]{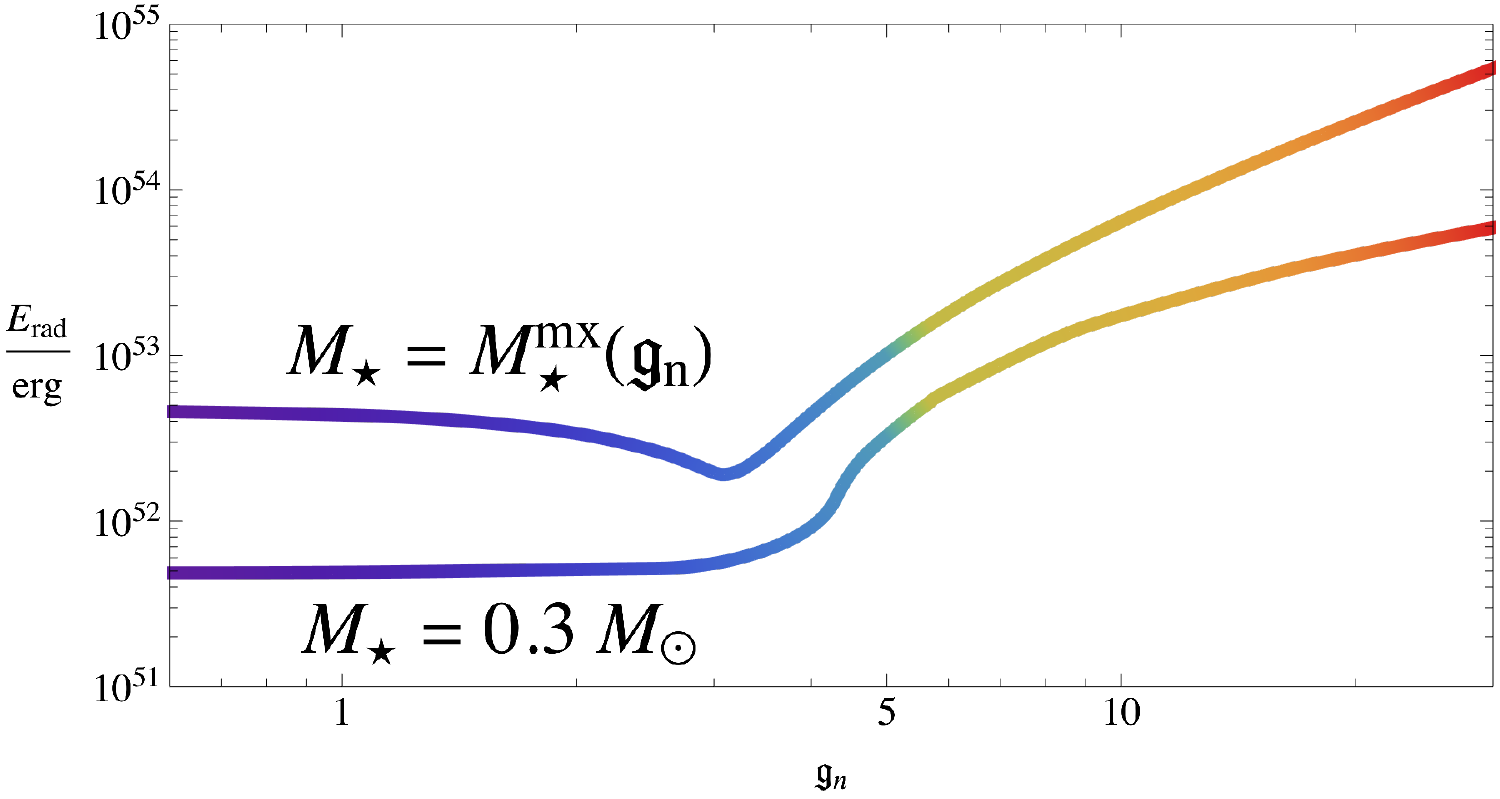}}
\caption{Plotted as a function of $g_\mathrm{n}\equiv\frac{g_\mathrm{n}\,m_\mathrm{n}}{m_\phi}$ for IGY neutron stars are the three quantities stated in the sub-captions. The lower curves are for stars with gravitational mass $M_\star=0.3\,M_\odot$ and the upper curves are for stars with the maximum mass  (see \S\ref{sec:MaximalStars})}
\label{fig:RestMassBindingEnergyRelease}
\end{figure}

Another important parameter of stellar remnants is the number of baryons they contain which we denote by $N_\star$.  Taking $m_\mathrm{n}$ as a conventional ``rest mass'' per baryon we can define the ``rest mass'' of a star by  $m_\mathrm{n}\,N_\star$.  This is plotted as a function of $\geff_\mathrm{n}$ in fig.~\ref{fig:RestMassBindingEnergyRelease}a for IGY neutron stars with gravitational mass $0.3\,M_\odot$ and for maximal stars.  One sees that the baryon content of a fixed gravitational mass stellar remnant increases for large $\geff_\mathrm{n}$ due to the increased binding energy due of the attractive $\phi$ force.  For strong coupling the maximal stars are hard stars which are predominantly bound by $\phi$ forces and can attain both large gravitational mass and large baryon content. 

One can define the average binding energy per baryon by $b_\star\equiv m_\mathrm{n}-M_\star/N_\star$ (using the conventional rest mass,$m_\mathrm{n}$  ).  This is plotted in fig.~\ref{fig:RestMassBindingEnergyRelease}b for $0.3\,M_\odot$ and maximal stars.  For large $\geff_\mathrm{n}$ the $0.3\,M_\odot$ (or any fixed $M_\star$) stars  approaches the hard drop limit: $b_\star\approx m_\mathrm{n}\,\left(1-\sqrt{\frac{\geff_\mathrm{vac}}{\geff_\mathrm{n}}}\right)$ where the gravitational binding energy becomes negligible (see \S\ref{sec:HardDrops}).  Maximal stars have only a slightly larger $b_\star$ due to non-negligible gravitational binding energy.

The large binding energy of hard stars is reflected in the energy liberated during their formation.  Denote the gravitational mass and baryon number of the progenitor star(s) by $M_0$ and $N_0$, respectively, so the binding energy per baryon of the progenitors is  $b_0\equiv m_\mathrm{n}-M_0/N_0$.  One can decompose the total energy released when the progenitor transitions to the remnant into the ``rest mass'' ejected, $M_\mathrm{ej}$, and the ``radiated'' energy, $E_\mathrm{rad}$.  These two components are given by\footnote{These equations slightly overestimate the actual rest mass ejected and underestimate the actual radiated energy because the ejecta is largely metals with rest mass per baryon a few MeV less than $m_\mathrm{n}$.}
\begin{eqnarray}
M_\mathrm{ej}&=&m_\mathrm{n}\,(N_0-N_\star)\\
E_\mathrm{rad}&=&M_0-M_\star-M_\mathrm{ej}=b_\star\,N_\star-b_0\,N_0
=\frac{b_\star}{m_\mathrm{n}-b_\star}\,M_\star-\frac{b_0}{m_\mathrm{n}-b_0}\,M_0 \ .\nonumber
\label{eq:ejecta}
\end{eqnarray}
Both $M_\mathrm{ej}$ and $E_\mathrm{rad}$ are potentially observable. The former, say, by observations of the SNe produced, the latter by measurements of neutrinos, gravitational radiation and photons emitted.  Neutrino emission constitutes the largest fraction of $E_{\rm rad}$.  A small outflow kinetic energy is also included in $E_{\rm rad}$ because much of it will be transformed to heating gas which will cool radiatively by photons. In most case the progenitor is a white dwarf or a nuclear burning star at the end of its lifetime in which case $b_0\,N_0\ll b_\star\,N_\star$ so $E_\mathrm{rad}\approx\frac{b_\star}{m_\mathrm{n}-b_\star}\,M_\star$.  This approximation for $E_\mathrm{rad}$ is plotted as a function of $\geff_\mathrm{n}$ in fig.~\ref{fig:RestMassBindingEnergyRelease}c.

From fig.~\ref{fig:RestMassBindingEnergyRelease}c figure one sees that $E_{\rm rad}$ increases rapidly with increasing $\geff_\mathrm{n}$.   For  $\geff_\mathrm{n}$ significantly greater than $\geff_\mathrm{vac}$ when the stellar structure is close to that of hard drops so $E_{\rm rad}\approx\left(\sqrt{\frac{\geff_\mathrm{n}}{\geff_\mathrm{vac}}}-1\right)\,M_\star$ and the maximum energy radiated by the maximal star is
\beq
E_\mathrm{rad}^\mathrm{mx} \approx\frac{\geff_\mathrm{n}}{12}\,\left(\sqrt{\frac{\geff_\mathrm{n}}{\geff_\mathrm{vac}}}-1\right)\,M_\odot=8.5\times10^{53}\,\mathrm{erg}\,\times\,\frac{\geff_\mathrm{n}}{\geff_\mathrm{vac}}\,\left(\sqrt{\frac{\geff_\mathrm{n}}{\geff_\mathrm{vac}}}-1\right)
\label{eq:EradMax}
 \eeq
for $\geff_\mathrm{n}>\geff_\mathrm{vac}$. The source of this extra energy is the large binding energy associated with the scalar mediated attraction between nucleons.  For these larger couplings the radiated energy greatly exceeds that produced by formation of a SM OV star.  Thus we find that Yukawa coupled neutron stars produce anomalously large amounts of radiated energy, mostly in the form of a neutrino burst, during their formation. Measuring neutrino bursts from a  star collapsing into a neutron star probes the scalar coupling of nucleons without direct observation of either the progenitor or remnant star.

\subsection{IGY Star Anomalies}
\label{sec:IGYStarAnomalies}

In \S\ref{sec:MassRadius}-\S\ref{sec:BindingEnergy} we describe properties of stars with the IGY EoS .  These stellar solutions have a number of qualitative and quantitative features which differ from what one finds in most if not all SM EoSs.  These are all related to the  EoS anomalies of \S\ref{sec:anomalies} though some appear at even weaker coupling than the EoS anomalies.  All of these anomalies involve observable quantities  which allow one to limit or detect scalar coupling of nucleons by astronomical observation of neutron stars.

Here we list several {\it mass/radius anomalies}:
\begin{itemize} 
\item {\bf mass versus radius inversion} : Stable hydrostatic solutions with SM EoSs have ${M_\star}'(R_\star)<0$.  This is not true for all stable IGY stars:  some hybrid stars with moderate coupling (fig.~\ref{fig:MofRmoderateIGY}) and some hard stars (all hard drops) with strong coupling (fig.~\ref{fig:MofRstrongIGY}) have ${M_\star}'(R_\star)>0$.  
\item {\bf maximum mass / minimum radius dichotomy} : For SM EoSs stable cold  hydrostatic configurations have a maximum mass, $M_\star^\mathrm{mx}$.  Since ${M_\star}'(R_\star)<0$ this {\it maximal star} also has the minimum radius.  In contrast for IGY stars with couplings $\geff_\mathrm{n}\gtrsim5.4$ the minimum radius configuration is not the maximal star (fig.s~\ref{fig:MofRmoderateIGY}~\&~\ref{fig:MofRstrongIGY}). 
\item {\bf radius gap} :  For SM EoSs there is a minimum radius below which there are no stable configurations and stable configurations extend from this minimal radius up to an instability gap between neutron stars and white dwarfs.  In contrast for IGY stars there is a gap in radii in the neutron star range ($\sim10$\,km) for $\geff_\mathrm{n}\gtrsim4$.
\item {\bf mass degeneracy} : For SM EoSs there is single configuration with a given mass and km scale radius.  In contrast for IGY stars there are are two stable solutions with different radii over some mass range if $\geff_\mathrm{n}\gtrsim4$.
\item {\bf small stars} :  SM EoSs have stable solutions with a minimum radius of several km.  In contrast hard stars with strong coupling (fig.~\ref{fig:MofRstrongIGY}) have stable solutions (hard drops) extending down to ``microscopic'' radii: $R_\star\sim{m_\phi}^{-1}$.
\item {\bf large stars} :  SM EoSs have stable solutions with a maximum mass of a few $M_\odot$ with a radius of $\sim10\,\mathrm{km}$.  In contrast hard stars with $\geff_\mathrm{n}>30$ have stable solutions with much larger masses and these maximal stars have much larger radii ($M_\star\approx\frac{\geff_\mathrm{n}}{12}\,M_\odot$, $R_\star\approx0.4\,\geff_\mathrm{n}\,\mathrm{km}$, see \S\ref{sec:MaximalStars} and fig.s~\ref{fig:MofRstrongIGY}~\&~\ref{fig:MRmaximalIGY}).
\end{itemize} 
As more neutron star radius measurements become available each of these anomalies will provide qualitative probe of nucleon Yukawa coupling.  An exception might be the small star / hard drops phenomena.  Whether hard drops would actually form in the course of normal stellar evolution is uncertain and even if they did their small mass and radius would make them difficult to detect.  If no such anomalies are found only weak Yukawa coupling will be allowed.  

Other anomalies in addition to those related to mass and radius anomalies include
\begin{itemize} 
\item {\bf thin crusts} : Stable hydrostatic solutions with SM EoSs contain superfluid neutrons with a crust of normal matter which is $\sim1\,$km thick.  While we have not addressed the issue of superfluidity in IGY we note that some hybrid stars (see \S\ref{sec:DensityProfiles} \& fig.~\ref{fig:DensityProfiles03SM}) and all hard stars (see \S\ref{sec:VacuumHardInterface} \& fig.~\ref{fig:VacuumBoundary}) are nearly uniform up to an extremely rapidly falling density near the surface.  Thus we would expect any crust to be very thin in these stars.  The outer region of a hard star is ``microscopic'' with thickness $\sim{m_\phi}^{-1}$. Pulsar glitches are believed to originate from rotational slippage between the superfluid and normal matter in the crust.   This crust must have sufficient moment of inertia to produce observed glitches which is unlikely to be the case in hard and some hybrid stars (see ref.~\cite{Caldwell1991} for this argument applied to strange stars).  Thus pulsar timing also places limits on scalar coupling to nucleons.
\item{\bf large neutrino bursts} : The energy radiated as a result of collapse to a SM neutron star is limited by the gravitational binding energy of the neutron star.  In contrast an IGY neutron star has additional binding energy due to the attractive $\phi$ mediated force between nucleons which will lead to much larger radiated energy if $\geff_\mathrm{n}\gtrsim10$: $E_\mathrm{rad}^\mathrm{mx}\approx8.5\times10^{53}\,\mathrm{erg}\,\times\,\frac{\geff_\mathrm{n}}{\geff_\mathrm{vac}}\,\left(\sqrt{\frac{\geff_\mathrm{n}}{\geff_\mathrm{vac}}}-1\right)$ (see \S\ref{sec:BindingEnergy} \& fig.~\ref{fig:RestMassBindingEnergyRelease}c).  This will lead to larger neutrinos bursts produced during collapse and cooling which would be observed by neutrino telescopes.
\end{itemize}
 
\section{Generality of IGY Phenomenology}
\label{sec:IGYgenerality}

While the IGY EoS is an idealization which does not provide a realistic model of matter throughout the entirety of any known astrophysical object the phenomenology it reveals can be considered generic in the sense that the unusual phenomenology it exhibits in stellar structure is purely a function of a simple feature in the EoS, i.e. in the $P(\ed)$ curve.  This feature is what one may call a {\it dip} in the EoS which softens with increasing density (as SM EoSs do) but then stiffens again.  In SM stars one finds such a dip between the densities where electrons become relativistic and where neutron pressure starts to dominate which leads to the radius gap between neutron stars and white dwarfs.  In IGY an effective Yukawa coupling, $\geff_\mathrm{n}\gtrsim4$ adds an additional dip of sufficient magnitude to cause a 2nd radius gap on the $R_\star\sim10\,$km scale which widens with larger $\geff_\mathrm{n}$.  For $\geff_\mathrm{n}\gtrsim5$ the dip is large enough so that $\frac{\partial P}{\partial\ed}<0$ in the dip which leads to a density gap between soft and hard matter allowing for hybrid stars which can have much smaller radii.  For $\geff_\mathrm{n}\gtrsim5.7$ the even larger dip has $P<0$, producing two classes of stars: hard and soft.  Hard stars have a minimum density and can be extremely small and are quite different from SM stars.

An EoS dip occurs naturally when SM particles are coupled to each other by a light scalar (force carrier).  The light mass of the scalar allows the superposition of the exterior $\phi$-field of many SM particles, the analog of the electric field of charged particles, to coherently add to a large $\phi$ polarization when the SM particle density is large. Stellar remnants with their large particle density is a natural place for this polarization to manifest.  The scalar polarization, $\tilde{\phi}$, will increase with density and is accompanied by increasing negative pressure, $-V(\tilde{\phi})$, and softening of $P(\ed)$.  As the SM particles become relativistic the $\phi$ polarization begins to saturate and so does  $-V(\tilde{\phi})$ so the EoS stiffens at higher densities.  This is a result of two effects: 1) the Yukawa coupled ``$\phi$ charge'' per particle is smaller for relativistic particles: $\tilde{n}=n\,\langle\gamma^{-1}\rangle$ from eq.~\ref{eq:n_free}; 2) the Yukawa coupling decreases the fermion kinematic mass as the polarization increases causing the particles to become relativistic at lower densities: $\tilde{m}_f=m_f-g_f\,\tilde{\phi}$ from eq.~\ref{eq:kinematicMass}.  This softening and then stiffening is the dip which leads to the phenomenology described in this paper.  It derives from the form of a Yukawa term in the Lagrangian.  Adding more realism to IGY should not change these basic features.

Since the baryonic chemical potential is $\mu_\mathrm{b}=(\varepsilon+P)/n_\mathrm{b}$ (eq.~\ref{eq:ChemicalPotential}) a dip in $P$ is associated with a dip in $\mu_\mathrm{b}$.  This decrease in $\mu_\mathrm{b}$ is due to tighter binding of the baryons caused by baryonic self-attraction mediated by $\phi$.  More tightly bound matter causes stars carrying the same baryon number to be smaller in size.  A larger dip corresponds to larger scalar binding energy, and if large enough one has stars which are primarily bound by the scalar force rather than by gravity as in hard stars.

\subsection{White Dwarfs}
\label{sec:WhiteDwarfs}

A somewhat trivial example of the generality of the IGY model comes from white dwarfs with a Yukawa coupling to the electrons (or more generally to leptons).  Since the pressure is primarily from degenerate free electrons and the density is primarily from non-relativisitic nuclei one has
\beq
\ed\approx n_\mathrm{b}\,m_\mathrm{n}+\ed_\mathrm{IGY} \qquad P\approx P_\mathrm{IGY}
\eeq
where one takes $f\rightarrow\mathrm{e}$ in the IGY EoS and the baryon number density is $n_\mathrm{b}=\sum_{(Z,A)}A\,n_{(Z,A)}$.  Here $Z$ is the atomic number and $A$ the atomic mass number and $n_{(Z,A)}$ is the number density of specific nuclei.  Charge neutrality requires $n_\mathrm{e}=\sum_{(Z,A)}Z\,n_{(Z,A)}$.  The majority white dwarfs have low mass and isotopically consist almost entirely of $^{12}_{\ 6}$C, $^{16}_{\ 8}$O, $^{20}_{10}$Ne, $^{24}_{12}$Mg each with $\frac{Z}{A}=\frac{1}{2}$.  In this case $n_\mathrm{b}\approx2\,n_\mathrm{e}$.  More massive white dwarfs with Fe cores have $\overline{Z/A}$ no more than 8\% lower.  Thus the IGY model gives an accurate pressure for a white dwarf but the energy dependence of $P(\ed)$ is rescaled relative to that in fig.~\ref{fig:EoSIGY1}; roughly shifted by a factor of $\sim4000$ when the electrons are non-relativistic. For fixed $n_\mathrm{e}/n_\mathrm{b}$ one can show that the values of $\geff_\mathrm{crit}$, $\geff_0$ and $\geff_\mathrm{vac}$ are unchanged by this rescaling.  So the dips in the EoS which lead to anomalous phenomenology occur at the same $\geff_\mathrm{e}$ as the $\geff_\mathrm{n}$ in the IGY neutron star model.  Thus one should expect very similar phenomenology for white dwarfs as for IGY neutron stars but of course masses and radii will differ numerically.  This will be worked out in a future work and compared to astronomical observations of white dwarfs.

\section{Amelioration of IGY Deficiencies}
\label{sec:IGYdeficiencies}

IGY is not realistic for neutron stars in a number of ways:
\begin{itemize}
\item IGY only models a single conserved fermion which we take to be the neutron for neutron stars.  This is only a rough approximation over a narrow range of densities since protons will carry a significant fraction of the baryon number especially at densities not much lower than nuclear densities.
\item The protons which are not present in IGY are accompanied by electrons and muons which dominate the pressure at low densities.  This leads to a more complicated $P(\ed)$ at low densities and $M_\star(R_\star)$ for small masses than is reflected in the IGY model.
\item IGY treats the neutrons as free particles with no interactions whereas it is known that nucleons experience significant nuclear interactions at densities above the saturation density which is at or below the density scale of neutron stars.  Thus IGY is certainly inaccurate for neutron stars just as is the OV model of neutron stars.
\end{itemize}
One can ameliorate some of these deficiencies with extensions of the IGY model.

Much of the formalism for an extending IGY to ameliorate these deficiencies have been given in this paper.  \S\ref{sec:freefermion}-\S\ref{sec:HeavyLightUltralight} include the energy density and pressure of multiple free fermionic species, with or without Yukawa couplings; as well as allowing for non-quadratic scalar potentials.  These generalizations were dispensed with in the sections that followed in order to focus on a simple, specific model with a semi-analytic EoS which exhibits qualitatively the EoS and stellar structure phenomenology of adding a Yukawa coupling.

Interactions among fermions were not considered here but are important for neutron stars where nuclear interactions play a role.  The form of $\ed_f$ and $P_f$ from IGY can be used only so long as the fermions move freely and not bound in localized structures such as quarks in nucleons or nucleons in a lattice but even so one must add the contribution of the interaction energy to $\ed$. The interaction contribution to $P$ is then given by the adiabaticity equation.  There is currently a great deal of uncertainty SM nuclear interactions and a model where light scalar is coupled to quarks and/or gluon further complicates modeling since the nuclear force carriers are themselves made of quarks and gluons.  Constructing a more realistic Yukawa coupled EoS for neutron stars will be left to future work. 


\section{Implications}
\label{sec:Implications}

A summary of the results of this paper has already been given in \S\ref{sec:Overview}.  Here we try to draw more general conclusions about the prospect for Yukawa coupled stars both for fitting data and excluding light scalars coupled to SM matter.  This paper gives the qualitative phenomenology of Yukawa coupled cold remnant stars with only rough quantitative predictions.  We have identified three mass regimes for scalars heavy, light and ultralight which depends on the type of star under consideration.  This paper focuses on the intermediate light scalar mass range which is valid throughout a neutron star if  $10^{-10}\,\mathrm{eV}\lesssim m_\phi\lesssim10^3\,\mathrm{eV}$ (eq.~\ref{eq:lightScalarMassRangeNS}) and throughout a white dwarf if $10^{-13}\,\mathrm{eV}\lesssim m_\phi\lesssim10^3\,\mathrm{eV}$ (eq.~\ref{eq:lightScalarMassRangeWD}).  In this mass range we find that for a star whose pressure is dominated by a single fermion species $f$ (neutrons for neutron stars and electrons for white dwarfs) the equation of state and thus the stellar structure depends on the effective coupling $\geff_f=\frac{g_f\,m_f}{m_\phi}$.  Thus this is only a one parameter family of stars resulting from the two parameters: the Yukawa coupling $g_f$ and the scalar mass $m_\phi$.  In our simple model various anomalies in the stellar structure (see \S\ref{sec:IGYStarAnomalies} and preceding sections) occur for $\geff_f\gtrsim3$.  The onset of anomalous stellar structure will occur for somewhat different values of $\geff_f$ for EoS models with greater realism but we expect correcting the inadequacies listed in \S\ref{sec:IGYdeficiencies} will only change these values by $\mathcal{O}(1)$.  If no anomalous behavior is observed then one should expect {\it very roughly} that this limits $\geff_f\lesssim\mathcal{O}(1)$ or $g_f\lesssim m_\phi/m_f$.\footnote{If $V(\phi)=\frac{1}{2}\,{m_\phi}^2\,\phi^2$ this bound on $g_\mathrm{n}$ is much weaker than laboratory bounds for all values of $m_\phi$ consistent with the LSA.  However with an appropriate choice of $V(\phi)$ one can make the effective scalar mass, $\tilde{m}_\phi$, at the nuclear densities of neutron stars much smaller than that of terrestrial laboratories; greatly weakening the laboratory bounds relative to the bound from stellar structure.}. To obtain much stronger bounds from stellar structure would require precision measurements of stellar parameters.

For large $\geff_f$ ($\ge 5.8$ in our simple model) the attractive binding force from the scalar field allows for a high density state of matter (hard matter) which is more tightly bound than free fermions and all low density matter is metastable to conversion to hard matter.  There is no direct path for conversion of low density matter so this instability would not manifest unless low density matter comes into contact with pre-existing seeds of hard matter.  It is conceivable that the dark matter is made of drops of hard matter, a possibility which will be explored in future work.

\section*{Acknowledgement}

We thank Gordan Baym, Reed Essick and Ivan Esteban for useful discussions as well as Paddy Fox for comments. CG is grateful to the generous support from Aspen Center for Physics (ACP) where part of the work was done. ACP is supported by National Science Foundation grant PHY-1607611.  AS and CG are supported by Fermilab, operated by the Fermi Research Alliance, LLC under Contract DE-AC02-07CH11359 with the U.S. Dept. of Energy.

\clearpage

\appendix
\section{$\langle\bar{\psi}\psi\rangle$ in Terms of Density and Pressure}
\label{app:psibarpsi}

Here we re-express $\langle\bar{\psi}\psi\rangle$ in terms of the density and pressure contributed by a free fermion species. We may ignore space-time curvature in solving $\langle\bar{\psi}\psi\rangle$, when a negligible fraction of particles have very low momentum: $p\,R\lesssim1$ (momentum $p$, curvature scale $R$).  This is true for all the SM fermions in a neutron star.  For flat space-time quantum field theory the Dirac field operator, $\psi$, is
\beq
\begin{split}
\psi         &= \int \frac{d^3{\bf p}}{(2\pi)^3}\frac1{\sqrt{2E}}
\sum_s\left(a_{\bf p}^s u^s({\bf p})e^{-i{ p\cdot x}}+{b_{\bf p}^s}^{\dagger}v^s({\bf p})e^{i{p\cdot x}}\right)\\
\bar{\psi}& = \int \frac{d^3{\bf p}}{(2\pi)^3}\frac1{\sqrt{2E}}
\sum_s\left({a_{\bf p}^s}^{\dagger} \bar{u}^s({\bf p})e^{i{ p\cdot x}}+{b_{\bf p}^s}\bar{v}^s({\bf p})e^{-i{p\cdot x}}\right)
\end{split}
\eeq
where $E=\sqrt{|{\bf p}|^2+m^2}$, $m$ is the particle mass, $a_{\bf p}^s$ (${a_{\bf p}^s}^{\dagger}$) annihilates (creates) a particle with momentum ${\bf p}$ and spin $s$, and $b_{\bf p}^s$ (${b_{\bf p}^s}^{\dagger}$) creates (annihilates) an anti-particles with momentum ${\bf p}$ and spin $s$. The creation and annihilation operators satisfy the anticommutation relations:
\beq
\{a^s_{\bf p},{a^{s'}_{\bf p'}}^{\dagger}\}=\{b^s_{\bf p},{b^{s'}_{\bf p'}}^{\dagger}\}=(2\pi)^3\delta^3({\bf p}-{\bf p'})\,\delta^{s\,s'}
\eeq
so
\beq
\begin{split}
\bar{\psi}\psi\supset&\sum_{ss'} \int \frac{d^3{\bf p}}{(2\pi)^3} \frac{d^3{\bf p'}}{(2\pi)^3}\frac{1}{2\sqrt{E\,E'}}
\times\\
&\left(
{a_{\bf p}^s}^{\dagger} a_{\bf p'}^{s'} 
\bar{u}^s({\bf p})u^{s'}({\bf p'})e^{-i{\bf(p- p')\cdot x}+i(E-E')t}
+{b_{\bf p}^s}{b_{\bf p'}^{s'}}^{\dagger}\bar{v}^s({\bf p})v^{s'}({\bf p'})e^{i{\bf(p- p')\cdot x}-i(E-E')t}\right)
\end{split}
\eeq

Apply $V^{-1}\int d V$ to $ \bar{\psi}\psi $,
\beq
\begin{split}
-\frac 1V\int dV \,\bar{\psi}\psi \sim & -\frac 1V
 \int \frac{d^3{\bf p}}{(2\pi)^3}\frac{1}{2E_{\bf p}}\sum_{ss'}\left(
{a_{\bf p}^s}^{\dagger} a_{\bf p}^{s'} 
\bar{u}^s({\bf p})u^{s'}({\bf p})
-{b_{\bf p}^{s'}}^{\dagger}{b_{\bf p}^s}\bar{v}^s({\bf p})v^{s'}({\bf p})\right)\\
=&  -\frac 1V
 \int \frac{d^3{\bf p}}{(2\pi)^3}\frac{2m}{2E_{\bf p}}\sum_{s}\left(
{a_{\bf p}^s}^{\dagger} a_{\bf p}^{s'} 
+{b_{\bf p}^{s'}}^{\dagger}{b_{\bf p}^s}\right)\\
= & -\frac 1{Vm}
 \int \frac{d^3{\bf p}}{(2\pi)^3}\left(E_{\bf p}-\frac{{\bf p}^2}{E_{\bf p}}\right)\sum_{s}\left(
{a_{\bf p}^s}^{\dagger} a_{\bf p}^{s'} 
+{b_{\bf p}^{s'}}^{\dagger}{b_{\bf p}^s}\right)
\end{split}
\eeq
where we have used $\bar{u}^s({\bf p}) u^{s'}({\bf p})=-\bar{v}^s({\bf p})v^{s'}({\bf p})=2m\delta^{ss'}$
in the second line.
Therefore, 
$\langle\bar{\psi}{\psi}\rangle$ approximately measures $(\ed -3P)/m$, assuming an ideal Fermi gas. 
In the non-relativistic limit, this is approximately equal to the number density of fermion $n$.

\section{TOV Equations with Anisotropic Pressure}\label{app:tov}
Consider a spherically symmetric metric, 
\beq\label{eq:metric}
ds^2=-e^{2\Phi}dt^2+\left(1-\frac {2Gm}{r}\right)^{-1}dr^2+r^2d\Omega^2
\eeq
where $\Phi=\Phi(r),m=m(r)$ under hydrostatic equilibrium. The Einstein tensor is given by 
\beq
G_{\mu\nu}\equiv R_{\mu\nu}-\frac12 Rg_{\mu\nu}
\eeq
In the basis of $(\hat{t},\hat{r},\hat{\theta},\hat{\phi})$,
\beq\label{eq:et}
\begin{split}
{G^t}_t&=-\frac{2Gm'}{r^2}\\
{G^r}_r&=
\frac{-2 Gm+2 r (r-2 Gm) \Phi'}{r^3}
\\
{G^{\theta}}_{\theta}&=\frac{(1+r\Phi')\left(Gm-Gm'r+r(r-2Gm)\Phi'\right)+r^2(r-2Gm)\Phi''}{r^3}
\\
{G^{\phi}}_{\phi}&={G^{\theta}}_{\theta}
\end{split}
\eeq
The Einstein equations are
\beq\label{eq:eins}
{G^{\mu}}_{\nu}=8\pi G {T^{\mu}}_{\nu}
\eeq

An important property of the stress-energy tensor is that it satisfies the continuity equation
\beq
\nabla_{\mu} {T^{\mu}}_{\nu}=0
\eeq
We consider a system that is static and spherically symmetric, described by the metric \eqref{eq:metric}.
Due to the symmetry, the $t$'s and $\phi$'s components of the continuity equation are automatically satisfied. 
The $\theta$'s component yields
$
{T^{\theta}}_{\theta}={T^{\phi}}_{\phi}.
$
The $r$'s component yields
\beq\label{eq:con}
0=\partial_r {T^r}_r+\Phi'\left( {T^r}_r- {T^t}_t \right)+
\frac 2r\left( {T^r}_r- {T^{\theta}}_{\theta} \right)
\eeq
If we identify $T^t_t=-\ed, T^r_r=P_{\parallel},T^{\theta}_{\theta}=P_{\perp}$, the continuity equation \eqref{eq:con} becomes
\beq\label{eq:con1}
0=\partial_r P_{\parallel}+\Phi'\left(P_{\parallel}+\ed\right)+
\frac 2r\left( P_{\parallel}- P_{\perp} \right)
\eeq

Now we can write down the TOV equations. Because of the continuity property of the energy stress tensor, not all components of Einstein equations are independent. Therefore, using \eqref{eq:con1},
the Einstein equations \eqref{eq:eins}  become
\beq
\begin{split}
\frac {dm}{dr}&=4\pi r^2 \ed \\
\frac{dP_{\parallel}}{dr}
& =-\frac{\ed Gm}{r^2} \left(\frac{ P_{\parallel}}{\ed}+1 \right)\left(\frac{4\pi  P_{\parallel}r^3}m+1\right) \left(1-\frac{2 Gm}r\right)^{-1}-\frac 2r\left( P_{\parallel}- P_{\perp} \right)
\end{split}
\eeq
In the case of a perfect fluid, $P_{\parallel}=P_{\perp}$, it simplifies to
\beq\label{eq:tov}
\begin{split}
\frac {dm}{dr}=&4\pi r^2 \ed \\
 \frac{dP}{dr}=&-\frac{\rho G m}{r^2} \left(\frac{4\pi P r^3}m+ 1\right)\left(\frac P{\ed}+1\right) \left(1-\frac{2Gm}r\right)^{-1}
\end{split}
\eeq

\section{Ideal Degenerate Fermi Gas}\label{app:ideal gas}

To solve TOV, one needs to find the relation between $\ed$ and $P_{\parallel},P_{\perp}$, a.k.a. the equation of state (EoS).
For an ideal fermi gas in equilibrium, the distribution function is
\beq
f(E)=\frac1{\textrm{exp}[(E-\mu)/kT]+1}
\eeq
For completely degenerate fermions ($T\to 0$), $\mu(T= 0)$ is called the Fermi energy $E_F$, and 
\beq
f(E)=\left\{
\begin{array}{cc}
1&E\le E_F\\
0&E> E_F
\end{array}\right .
\eeq
Define Fermi momentum $p_F$ as
\beq
E_F\equiv \left(p_F^2+m^2\right)^{1/2}
\eeq

The number density of the degenerate ideal fermi gas is 
\beq\label{eq:n}
n= 2\int_0^{p_F} 4\pi p^2 dp =\frac1{3\pi^2} m^3x^3
\eeq
where we use the ``relativity parameter", $x$, define as
\beq\label{eq:pF}
x\equiv p_F/m
\eeq

The energy density is 
\beq\label{eq:E}
\begin{split}
\ed=2\int _0^{p_F}\left(p^2+m^2\right)^{1/2}4\pi p^2 dp = m^4 \xi(x)\\
\xi(x)=\frac1{8\pi^2}\left\{
x(1+x^2)^{1/2}(1+2x^2)-\ln\left[x+(1+x^2)^{1/2}\right]
\right\}
\end{split}
\eeq

The pressure is 
\beq\label{eq:P}
\begin{split}
P=\frac13 2\int _0^{p_F}\frac{p^2}{\left(p^2+m^2\right)^{1/2}}4\pi p^2 dp =m^4\varphi(x)
\\
\varphi(x)=\frac1{8\pi^2}\left\{
x(1+x^2)^{1/2}(2x^2/3-1)+\ln\left[x+(1+x^2)^{1/2}\right]
\right\}
\end{split}
\eeq
Clearly, by scanning over x, one can numerically obtain an EoS $\ed=\ed(P)$.

\section{Simpler Analytics for IGY}
\label{app:IGYsimplified}

An algebraically and numerically simpler form of the IGY EoS of eq.s~\ref{eq:EoS_IGY} is given by
\begin{align}
\label{eq:EoS_IGY_simple}
\frac{\ed_\mathrm{IGY}(x)}{m_f^4}&=\mathcal{A}(x)+\mathcal{B}(x) &
\frac{n_f(x)             }{m_f^3}&=\mr(x)^3\,\frac{x^3}{3\,\pi^2}           &
 \mathcal{A}(x)&=\mr(x)^4\,\frac{x^3\sqrt{1+x^2}}{4\pi^2} \nonumber \\
\frac{              P_\mathrm{IGY}(x)}{m_f^4}&=\frac{\mathcal{A}(x)}{3}-\mathcal{B}(x) &
\frac{ \tilde{n}_f(x) }{m_f^3}&=\frac{1-\mr(x)}{{\geff_f}^2} &
 \mathcal{B}(x)&=\frac{\left(1-\mr(x)\right)^2}{{\geff_f}^4}+\frac{\mr(x)\,(1-\mr(x))}{4}
\end{align}
where $\mr(x)$ is given in eq.~\ref{eq:muplus}.  When combined with $\mr'(x)=-\frac{{\geff_f}^2}{\pi^2}\,\frac{x^2}{\sqrt{1+x^2}}\,\frac{\mr(x)^4}{3-2\,\mr(x)}$ almost any quantity of interest can be expressed as a multinomial or ratio of multinomials in $\mr(x)$, $x$ and $\sqrt{1+x^2}$.  It is easier to derive many IGY identities, e.g. the adiabaticity equation, using these expressions.

\pagebreak

\bibliographystyle{unsrt}
\bibliography{main}

\end{document}